\documentclass[useAMS,usenatbib]{mn2e}
\usepackage{graphicx}
\usepackage{subfigure}
\usepackage{amsmath}

\title[Triggered star formation diagnostics]{Testing diagnostics of triggered star formation}
\author[T. J. Haworth, T. J. Harries and D. M. Acreman]{Thomas J. Haworth\thanks{E-mail:
haworth@astro.ex.ac.uk}, Tim J. Harries and David M. Acreman\\
School of Physics, University of Exeter, Stocker Road, Exeter EX4 4QL}
\begin{document}

\date{Accepted ???. Received ???; in original form ???}

\pagerange{\pageref{firstpage}--\pageref{lastpage}} \pubyear{2012}

\maketitle

\label{firstpage}

\begin{abstract}
We produce synthetic images and SEDs from radiation hydrodynamical simulations of radiatively driven implosion. The imaged bright rimmed clouds (BRCs) are morphologically similar to those actually observed in star forming regions. Using nebular diagnostic optical collisional line ratios, simulated Very Large Array (VLA) radio images, H$\alpha$ imaging and SED fitting we compute the neutral cloud and ionized boundary layer gas densities and temperatures and perform a virial stability analysis for each cloud. 
We determine that the neutral cloud temperatures derived by SED fitting are hotter than the dominant neutral cloud temperature by $1-2$\,K due to emission from warm dust. This translates into a change in the calculated cloud mass by $8-35$\,\%. Using a constant mass conversion factor ($C_{\nu}$) for BRCs of different class is found to give rise to errors in the cloud mass of up to a factor of 3.6.
The ionized boundary layer (IBL) electron temperature calculated using diagnostic line ratios is more accurate than assuming the canonical value adopted for radio diagnostics of $10^4$\,K. Both radio diagnostics and diagnostic line ratios are found to underestimate the electron density in the IBL. 
Each system is qualitatively correctly found to be in a state in which the pressure in the ionized boundary layer is greater than the supporting cloud pressure, implying that the objects are being compressed.
We find that observationally derived mass loss estimates agree with those on the simulation grid and introduce the concept of using the mass loss flux to give an indication of the relative strength of photo-evaporative flow between clouds. The effect of beam size on these diagnostics in radio observations is found to be a mixing of the bright rim and ambient cloud and HII region fluxes, which leads to an underestimate of the cloud properties relative to a control diagnostic.

\end{abstract}


\section{Introduction}
A central problem in the study of star formation and galaxy evolution is the effect of stellar feedback on star formation rates, efficiencies, and the initial mass function (IMF) \citep[e.g.][and references therein]{2011MNRAS.417.1318D, 2011EAS....51...45E,2012MNRAS.419.3115B, 2012arXiv1201.3659E, 2012arXiv1204.3552K}. Stellar winds drive into the surrounding material which is expected to either induce star formation by triggering collapse in otherwise stable conglomerations or to hinder star formation by dispersing material and driving turbulence \citep[example theoretical, observational and synthetic observational studies include][]{2001ASPC..243..757M, 2005MNRAS.357.1370S, 2010ApJ...715.1170A, 2011ApJ...743...91O}. Supernova feedback is another kinematic mechanism whereby the explosions of massive stars locally disperse potential star forming material, but may cause large-scale collapse of nearby gas clouds \citep[see][and references therein]{2011IAUS..270..309H}. Young stars, specifically the most massive (i.e. OB), emit large amounts of ionizing radiation into the surrounding gas. This quickly ionizes and heats the surroundings, generating shocks that sweep up material into a shell that can either become locally gravitationally unstable \citep[the collect and collapse scenario -][]{2005A&A...433..565D,2006A&A...446..171Z, 2007MNRAS.375.1291D} or drive into pre-existing density structures and compress them to form stars, this final process is known as radiatively driven implosion (RDI). 

RDI has been subject to a large amount of numerical modelling which has repeatedly demonstrated that the compression and implosion of clouds is possible \citep[e.g.][]{1982ApJ...260..183S,1989ApJ...346..735B,1994A&A...289..559L,2003MNRAS.338..545K,2009ApJ...692..382M,2009MNRAS.393...21G,2011ApJ...736..142B,2012MNRAS.tmp.2723D,2012MNRAS.420..562H}. The resulting objects are usually bow shaped thin dense shells or cometary structures, in qualitative agreement with studies such as the northern hemisphere bright rimmed cloud (BRC) survey of \cite{1991ApJS...77...59S}, which found that BRCs can be broadly classified by their rim morphology, being either: type A (moderately curved), type B (tightly curved) or type C (cometary). 

A number of bright rimmed clouds have been identified in the vicinity of which are often collections of young stars \citep[studies include][]{1991ApJS...77...59S,2002AJ....123.2597O,2004A&A...414.1017T, 2007ApJ...657..884L, 2009A&A...504...97B,2009MNRAS.400.1726M, 2009MNRAS.396..964C, 2010ApJ...717.1067C, 2011MNRAS.415.1202C}, these are often cited as examples of triggered star formation via RDI and have been studied using a range of techniques.
One of the most popular signatures of triggered star formation is an age gradient in YSOs, with bluer and therefore supposedly older sources located in closer proximity to the triggering star. \cite{2009A&A...504...97B} study the BRC IC 1396N in the Cep OB2 association and caution against this approach, as a collection of stars of similar age may exhibit an apparent observational age gradient if they are sequentially exposed to high ionizing flux and stripped of their circumstellar environment.

An alternative to studying stellar age gradients is to analyze the neutral cloud and ambient ionized gas properties. \cite{1994A&A...289..559L} calculated semi-analytic two dimensional models of RDI and included figures of the system emissivity which were found to have similar morphology to the BRC types observed in star forming regions. They further compared the cloud and ionized boundary layer (IBL) pressures using a multiwavelength molecular line and radio emission study of CG5 in IC1848 \citep{1997A&A...324..249L} and found greater pressures in the IBL by approximately a factor of 10, indicative of shock compression. They also calculated an estimate of the incident ionizing radiation flux and mass loss rates, finding values of $4.8\times10^9$\,cm$^{-2}$\,s$^{-1}$ and 105\,M$_{\odot}$\,Myr$^{-1}$ respectively. \cite{2004A&A...414.1017T} and \cite{2004A&A...426..535M} performed a similar analysis on a larger number of clouds, finding that for only a small number the IBL is at higher pressure and that more frequently the objects are in pressure balance. The aforementioned studies suggest that greater external pressure will lead to the eventual dispersal of the cloud, whereas clouds that are overpressured relative to the surroundings will stall the shock until enough material is accumulated and the pressure sufficiently increased to continue driving into the cloud. The presence of photo-evaporative flow in \cite{1994A&A...289..559L} and \cite{2004A&A...414.1017T} is also an important factor in identifying RDI. In addition to calculating the mass loss rate, striations in H$\alpha$ imaging about the BRC is identified as an indication of such a flow. As yet, there is no direct translation between numerical models and these observational diagnostics of BRCs.

In this paper we use the final state of the RDI models generated in \cite{2012MNRAS.420..562H} that incorporated the diffuse field. These calculations considered the RDI of a Bonnor-Ebert sphere (BES) at three distances from the triggering star and gave rise to clouds that appear to be type A (the low flux model), B (the high flux model) and type B-C (the medium flux model). These models also vary in the nature of the driving shock, with the medium and low flux models being driven by a photoionizing shock/photo-evaporative flow and the high flux model in pressure balance. With these final grid states we perform synthetic H$\alpha$ and radio continuum imaging, as well as generate synthetic SEDs to perform standard diagnostics that look for RDI. We also explore the use of long slit spectroscopy to calculate diagnostic forbidden line ratios, determine the nebular conditions and infer the stability of the BRCs. In addition we investigate the observational characteristics of photo-evaporative flows and mass loss rate calculations. This is all performed from the unique perspective in which the actual conditions of the system are known. Through this process we aim to guide observers towards unambiguous signatures of triggered star formation in BRCs and to investigate the accuracy and applicability of these standard diagnostic techniques.

\section{Numerical Method}

\subsection{Overview}
We used the grid based radiation transport and hydrodynamics code \textsc{torus} \citep{2000MNRAS.315..722H, 2010MNRAS.406.1460A, 2011MNRAS.416.1500H, 2012MNRAS.420..562H} to perform the calculations in this paper. Using the final states of the models which included the diffuse field in \cite{2012MNRAS.420..562H} we perform photoionization calculations with additional atomic species. Dust is then added to the resulting grids as part of post processing and the final grid states are used to generate SEDs and images for diagnostics. 

The models from \cite{2012MNRAS.420..562H} comprised a star at varying distances from a BES. Radiation from the star (which was off the edge of the grid) entered the grid at the $-x$ boundary in a plane parallel manner, under the assumption that the grid was sufficiently far from the ionizing star for this to be the case. An ionization front formed and accumulated material that was driven into the BES, compressing it. Material was allowed to stream freely off the edge of the $\pm\,x$ boundaries but not allowed to re-enter the grid. The other boundaries were periodic.

\subsection{Photoionization}
\label{photoion}
The photoionization routine follows that detailed in  \cite{2003MNRAS.340.1136E}, \cite{2004MNRAS.348.1337W} and \cite{2012MNRAS.420..562H} and involves propagating photon packets from sources through a computational grid. A photon packet is a collection of photons for which the total energy is constant, being the total luminosity of the stellar photon sources divided by the total number of packets. The number of photons in the packet varies with frequency. As photon packets traverse cells on the grid they modify the energy density in the cell region, allowing the ionization and temperature states to be calculated. The diffuse field is included in models in this paper, whereby a new photon packet is immediately generated upon absorption with a new frequency and direction and is further propagated over the computational domain. This process repeats until the packet escapes the grid. We also include periodic photon packet boundary conditions, without which the ionizing flux at the edges of the grid can be underestimated if the model geometry is periodic. However, in order to avoid long loops for low frequency packets which have a small probability of interaction, each photon packet is only allowed to cross a periodic boundary once. 
Ionization balance in all calculations is determined by solving the ionization balance equation \citep{1989agna.book.....O} and photoionization equilibrium is assumed. 
In this paper we include treatment of a range of atomic constituents: hydrogen, helium, carbon, nitrogen, oxygen, neon and sulphur.  Thermal balance is calculated by finding the temperature at which the heating and cooling rates match in the manner described in \cite{2004MNRAS.348.1337W}. Hydrogen and helium heating is treated as well as recombination, free-free and collisional line cooling.

\subsection{Synthetic Imaging and SEDs}
\label{imgseds}
Synthetic images are generated by accumulating photon packets from the grid in a 2D array of collecting bins. The basic method follows the Monte Carlo scheme detailed in \cite{1991A&A...247..455H} and \cite{2000MNRAS.315..722H}, an example application is \cite{2004MNRAS.351.1134K}. For an observer location (which can be specified arbitrarily) relative to the grid, a pixel array is generated. The emissivity across the grid is then calculated and can include dust continuum, free-free continuum, forbidden and recombination line radiation.
The image is constructed by propagation of photon packets from dust, gaseous and stellar sources, with each packet carrying power $P$ given by
\begin{equation}
	P = \frac{L_{*} + W_{\rm{tot}}}{N_{\rm{monte}}}
\end{equation}
where $L_*$, $W_{\rm{tot}}$ and $N_{\rm{monte}}$ are the total stellar source emission, the total emission from the gas and dust and the number of photon packets used in the image generation respectively. 

The Monte Carlo method is subject to Poisson noise analogous to that inherent in the collection of photons in real imaging, therefore we employ variance reduction techniques to reduce the calculation time required to reach adequate signal to noise. To improve sampling in the gas the probability of a photon packet originating from a stellar source is fixed at 0.1 and the packet assigned an appropriate weight $w$. Following initial generation and each scattering event of photon packets, an additional packet is forcibly directed towards the observer with a modified weighting the, so called, `peel-off' technique \citep{cashwell1959practical, 1984ApJ...278..186Y}. Once a photon packet escapes the grid in the direction of the observer it changes the flux $F_{\gamma}$ in whichever pixel it intersects to
\begin{equation}
	F_{\gamma + 1} = F_{\gamma} + \frac{P w \rm{e}^{-\tau}}{4\pi}
\end{equation}
where $w$ is the photon packet weight and $\tau$ is the optical depth along the packet's path to the observer. The accumulated photon packet contributions are converted into a final image in distance independent units of mega Janskys per steradian.

Spectral energy distributions are also generated via Monte Carlo radiative transfer in a similar manner. Rather than contributing to a 2D pixel array photon packet intensities are binned by frequency to form the SED \citep[see][]{2004MNRAS.351.1134K}. 

\subsubsection{Dust properties}
\begin{figure}
	\includegraphics[width=9cm]{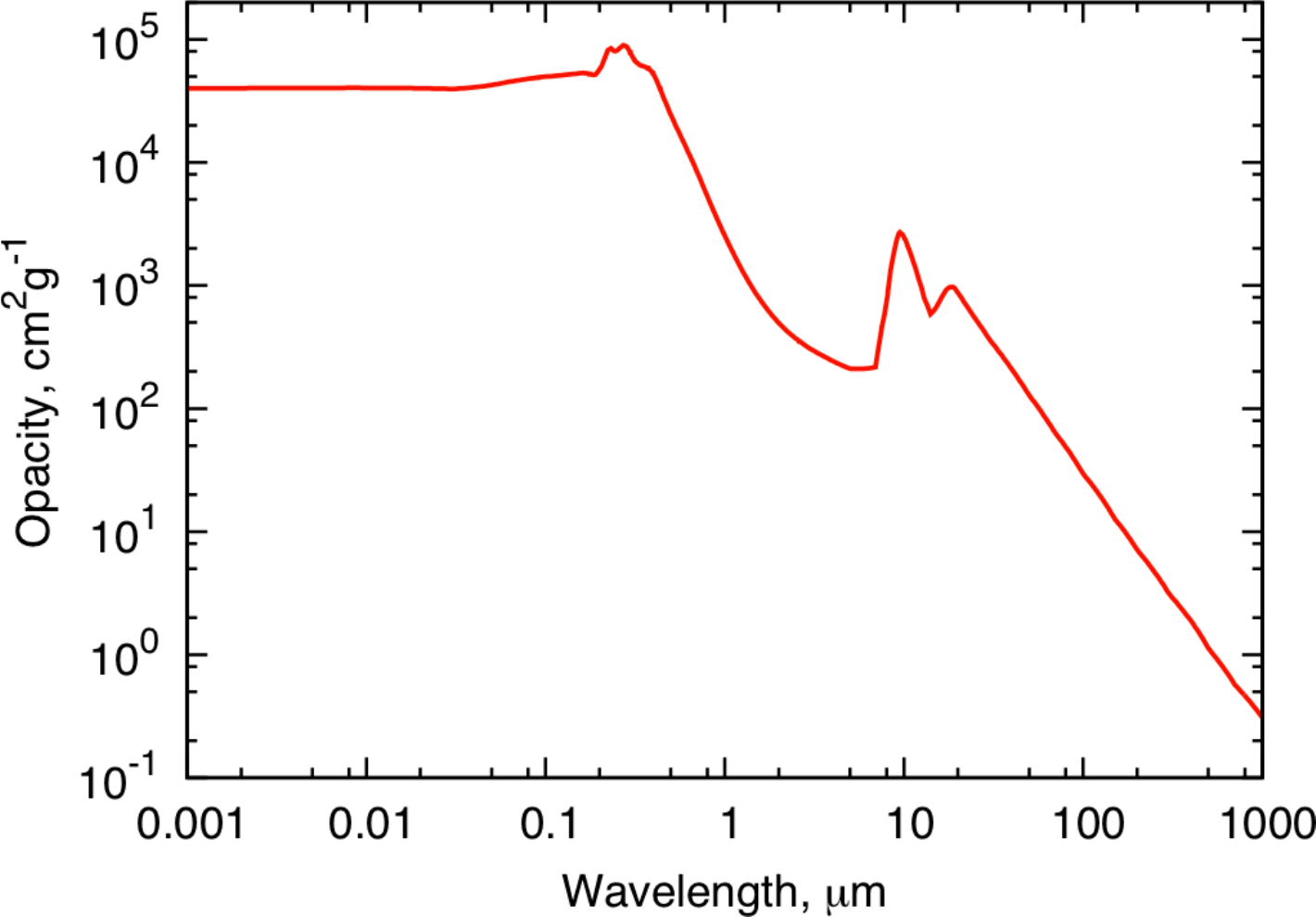}
	\caption{The dust opacity as a function of wavelength.}
	\label{opacity}
\end{figure}

Dust is not directly included in the photoionization calculations in this paper, rather it is added to all cells where the temperature is less than 1500\,K  prior to image/SED generation \citep[e.g.][]{2001ApJ...560..957D}. Unless otherwise specified, a dust to gas ratio of $1\times10^{-2}$ is used in all images generated in this paper. 
We assume spherical silicate dust particles that follow a standard interstellar medium power-law size distribution \citep[e.g.][]{1977ApJ...217..425M}. The optical constants are taken from \cite{1984ApJ...285...89D}. We use a pre-tabulated Mie-scattering phase matrix. The wavelength dependency of the dust opacity is given in Figure \ref{opacity}. Dust was not included in the original radiation hydrodynamic models because a simplified thermal balance calculation was used to remain consistent with previous models. A realistic treatment of the dust would therefore not be achieved. It is expected that dust would not significantly modify the radiation hydrodynamic calculations as it would be destroyed in the ionized medium, not altering the pressure external to the cloud and not increasing the supporting cloud pressure. However confirming this is a subject for future study.

\subsection{Radio Imaging}
\label{radImg}
Synthetic images generated at radio wavelengths using the method described in section \ref{imgseds} will have a sub-pixel beam and a Poisson noise level determined by the number of photon packets used in the image generation, at a level that is possibly lower than the Gaussian noise associated with radio imaging. We therefore modify images at these wavelengths as part of post processing. The image is smoothed to a Gaussian beam using \textsc{aconvolve} from \textsc{ciao} v4.1 \citep{2006SPIE.6270E..60FB} to a size appropriate to the half power beamwidth (HPBW) of the simulated instrument.
We assume that the Gaussian smoothing outlined above extinguishes the effects of Poisson noise (which is relatively small) and then add Gaussian noise to the image separately using the \textsc{starlink} \textsc{addnoise} routine. The emulated exposure time, beam size and noise level are chosen to be typical of those for a selected instrument.

\section{Diagnostic Techniques}
To determine what an observer would infer following observations of our model cloud we employ a range of standard diagnostic techniques that are used to determine the ionized and neutral gas properties. 

\subsection{Neutral cloud properties from SED fits}
\label{greyFitting}
In this work no molecular lines are treated. We therefore obtain the neutral cloud gas properties by fitting the submillimetre thermal dust continuum tail of the object SED with a greybody function of the form
\begin{equation}
	F_{\nu} = \Omega B_{\nu}(T_{\rm{d}})\left(1 - \rm{e}^{-\tau_{\nu}}\right)
	\label{greybody}
\end{equation}
where $\Omega$, $B_{\nu}(T_{\rm{d}})$ and $\tau_{\nu}$ are the solid angle subtended by the region for which the temperature is being derived, the frequency specific Planck function at dust temperature $T_{\rm{d}}$ and the optical depth at frequency $\nu$ respectively \citep{1983QJRAS..24..267H,1998MNRAS.301.1049D,2004A&A...414.1017T,2008A&A...477..557M}. We adopt a procedure in which the optical depth at a given frequency is paramaterised in terms of a reference frequency and optical depth i.e. 
\begin{equation}
	\tau_{\nu} = \tau_{\rm{ref}}\left(\frac{\nu}{\nu_{\rm{ref}}} \right)^{\beta}
	\label{tauGen}
\end{equation}	
\citep[e.g.][]{1983QJRAS..24..267H}. We choose a reference wavelength of 850\,$\mu$m and adopt a value of 2 for $\beta$, the index specifying the frequency dependency of the dust emissivity, following \cite{1983QJRAS..24..267H}, \cite{2004A&A...414.1017T} and \cite{2008A&A...477..557M}. The reference optical depth is calculated using the equation for submillimetre optical depth from \cite{1983QJRAS..24..267H}

\begin{equation}
	\tau_{\nu} = F_{\nu}\left[\pi \theta_{\rm{R}}^2 B_{\nu}(T_{\rm{d}}) \right]^{-1}
	\label{getTau}
\end{equation}
where $\theta_{\rm{R}}$ is the angular radius of the region over which the flux is being integrated in radians.
We fit the SED using a chi-square minimization.
Although we can fit across an entire calculated spectrum, we typically do so over only a small range in the manner of observational studies such as \cite{2004A&A...414.1017T}.
The cloud mass is then found using the established method of \cite{1983QJRAS..24..267H}, whereby the total gas and dust mass of the cloud is given by
%\begin{equation}
%	M = \frac{d^2 F_{\nu}C_{\nu}}{B_{\nu}(T_d)}
%%\end{equation}
%or
\begin{equation}
	M = \frac{d^2 F_{\nu}C_{\nu}}{B_{\nu}(T_{\rm{d}})}
	\label{dustGasMass}
\end{equation}
where $d$ is the distance of the cloud from the observer and $C_{\nu}$ is a mass conversion factor. A value for $C_{\nu}$ needs to be appropriately selected \citep{1983QJRAS..24..267H,1984ApJ...285...89D,1994A&A...291..943O, 2001ApJ...552..601K}. Typically, smaller values of $C_{\nu}$ apply to cold, high density, regions and larger values to low density regions (e.g. the diffuse ISM).  As an example, the value selected by \cite{2004A&A...414.1017T} for 850\,$\mu$m analysis of clouds with assumed densities of 10$^5$\,cm$^{-3}$ is 50\,g\,cm$^{-2}$. 
An expression for $C_{\nu}$ is given by \cite{1983QJRAS..24..267H}
\begin{equation}
	C_{\nu} = \left[N_{\rm{H}}/\tau_{\nu} \right]m_{\rm{H}}\mu
	\label{cnu}
\end{equation}
where $N_{\rm{H}}$ is the column density, $m_{\rm{H}}$ the hydrogen mass and $\mu$ is the mean particle mass relative to hydrogen, which we assume to be 1.36 in neutral gas following, e.g. \cite{1983QJRAS..24..267H,2006ApJ...650..933B}.

The model clouds are not supported by turbulent motions owing to the ideal starting conditions. We assume that the sound speed in the neutral gas is isothermal and is calculated using the fitted dust temperature.

\subsection{Ionizing flux, mass loss rates and IBL electron densities from radio emission}
\label{radio}
We calculate the conditions in the IBL of the BRC following the standard radio diagnostics of \cite{1997A&A...324..249L} which are used in, for example, \cite{2004A&A...426..535M}, \cite{2004A&A...414.1017T} and \cite{2006A&A...450..625U}. This technique does not constrain the temperature in the IBL so we assume the standard value of $10^4$\,K following the aforementioned studies.

An observational estimate of the ionizing flux $\Phi$ per square centimetre per second impinging upon the BRC can be made from the 20\,cm free-free emission integrated flux from the IBL by rearranging equation 6 from \cite{1997A&A...324..249L}
\begin{equation}
%	\begin{array}{c}
	\Phi = 1.24 \times 10^{10} F_{\nu}T_{\rm{e}}^{0.35}\nu^{0.1}\theta^{-2}
%\Omega
%	\left(\frac{\Phi}{\rm{cm}^{-2}\,\rm{s}^{-1}}\right) = \frac{10^9}{0.321}\left(\frac{F_{\nu}}{1\,mJy}\right)\left(\frac{T_e}{10^4\,K}\right)^{0.35}\left(\frac{\nu}{1\,GHz}\right)^{0.1} \\
%	\times\left(\frac{\theta}{10\arcsec}\right)^{-2} \\
%	\end{array}
	\label{ionizingFlux}
\end{equation}
where $T_{\rm{e}}$ is the electron temperature of the ionized gas in K, $\theta$ is the angular diameter in arcseconds of the region over which the flux is integrated and $F_{\nu}$ is the integrated flux in mJy at frequency $\nu$ in GHz. 
%This is also used by  \cite{2004A&A...426..535M} and \cite{2004A&A...414.1017T}.

The electron density can also be calculated using 
\begin{equation}
	n_{\rm{e}} = 122.41\sqrt{\frac{F_{\nu}T_{\rm{e}}^{0.35}\nu^{0.1}\theta^{-2}}{\eta R}}
	\label{radioNe}
\end{equation}
where $\eta$ is the thickness of the ionized boundary layer as a fraction of the cloud radius and $R$ is the cloud radius in parsecs \citep{1997A&A...324..249L}. \cite{1989ApJ...346..735B} give $\eta$ typically $\approx 0.2$ which is the value adopted by \cite{2004A&A...426..535M}, \cite{2004A&A...414.1017T} and \cite{2006A&A...450..625U}, we therefore also assume this value.  The electron density allows the ionized boundary layer gas pressure to be derived using
\begin{equation}
	P_{\rm{i}}= 2\rho_{\rm{i}} c_{\rm{i}}^2
	\label{pressureEqn}
\end{equation}
where $P_{\rm{i}}$, $\rho_{\rm{i}}$ and $c_{\rm{i}}$ are the pressure, density and sound speed in the ionized boundary layer \citep[e.g.][]{2004A&A...426..535M,2004A&A...414.1017T}.  We assume that in the ionized regions where these diagnostics are applied that $\rho_{\rm{i}} = n_{\rm{e}} m_{\rm{H}}$ and that the sound speed is isothermal, i.e. $c_{\rm{i}} = \sqrt{k_{\rm{B}}T_{\rm{e}}/\mu m_{\rm{H}}}$, where $k_{\rm{B}}$ is the Boltzmann constant and $\mu$ is the mean particle mass relative to hydrogen, here assumed to be 0.6 in the IBL, the value for ionized gas of solar composition.

The mass loss rate from the cloud due to ionized gas streaming away from the surface (photo-evaporative flow) is calculated using
\begin{equation}
	\dot{M} = 4.4\times10^{-3}\Phi^{1/2}R^{3/2}\rm{M}_{\odot}\rm{Myr}^{-1}
	\label{massLoss}
\end{equation}
where $R$ is the cloud radius in parsecs \citep{1994A&A...289..559L,2004A&A...414.1017T}. The strong photo-evaporative flows identified in some of the RDI model clouds from \cite{2012MNRAS.420..562H}, which are also studied here, will provide a valuable test of the mass-loss rate estimate from Equation \ref{massLoss}. 

\subsection{H\,$\alpha$ emission}
We generate images of H\,$\alpha$ recombination emission as it has previously been used to identify photo-evaporative flow via striations perpendicular to the bright rim \citep{2004A&A...414.1017T}. Imaging at this wavelength (6563\,\AA) is also useful for tracing ionized gas and is regularly used as a basis for identifying regions over which to perform long slit spectroscopy \citep[e.g.][]{2003A&A...412...69T}.

\label{lineRatioDescription}
\subsection{IBL properties from optical collisional line ratios}
\label{ratdesc}
The conditions of HII regions and planetary nebulae have long been studied using ratios of forbidden lines that are sensitive to the electron density or temperature \citep[e.g.][and references therein]{1989agna.book.....O,1998ppim.book.....S, 2000MNRAS.311..317C, 2000MNRAS.311..329D, 2012MNRAS.420.2280L}. Single forbidden lines have been used in studies to identify pre main sequence (PMS) stars in and around BRCs as an indicator of triggered star formation \citep[e.g.][]{2005ApJ...624..808L, 2007ApJ...657..884L}. However, the line ratios have not yet been applied to calculating the IBL properties and hence the relative pressures of the IBL and the neutral cloud. We therefore derive the ionized gas conditions using these diagnostic ratios, in addition to the radio analysis, to assess their future use as a tool for identifying RDI. 

Forbidden line intensities are usually obtained by performing slit spectroscopy of the system \citep[e.g.][]{2003A&A...412...69T}. We emulate this form of analysis by generating synthetic images at a range of forbidden line wavelengths and calculating the intensity across a pseudo-slit region on the image. This is akin to choosing the intensity at a specific wavelength on the slit spectrum.

\begin{figure}
	\hspace{-13.5pt}
	\includegraphics[width=9cm]{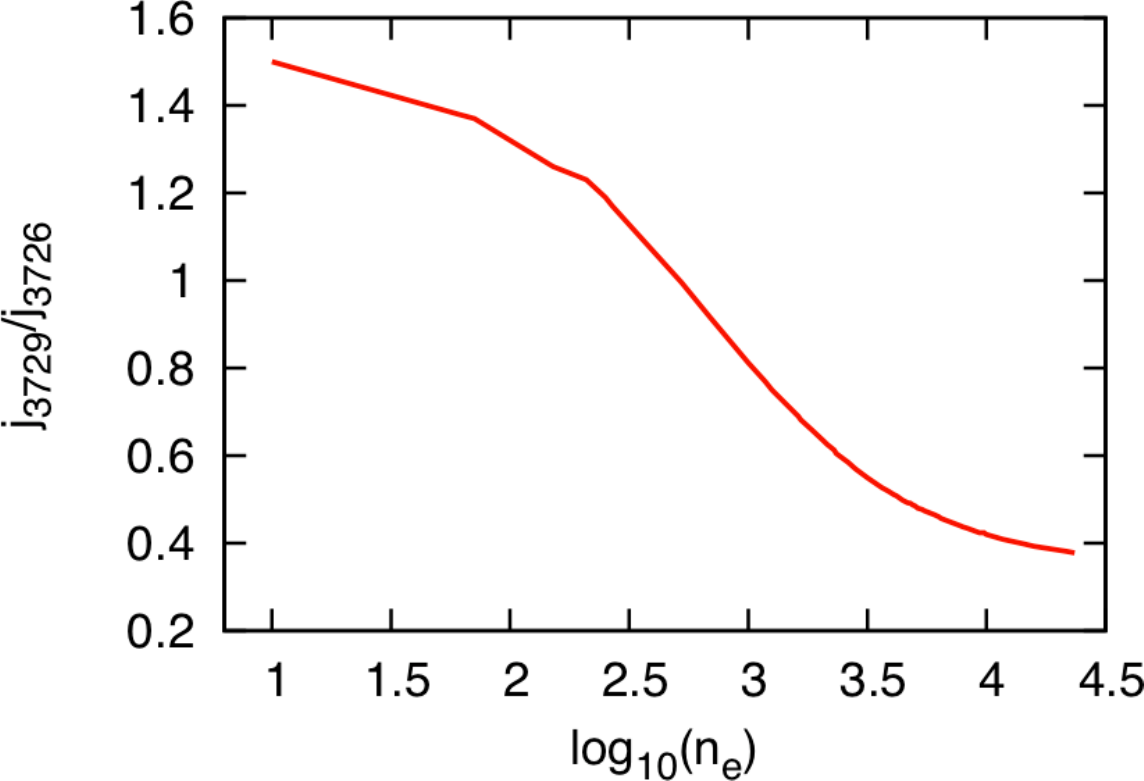}
	\caption{The [O II] (3729\,\AA/3726\,\AA) ratio variation with logarithmic electron density.}
	\label{neRats}
\end{figure}

We calculate the electron density using  the [O II] 3729\AA/3726\AA\, ratio by linearly interpolating between the tabulated ratio and electron density values given in Table 2 of \cite{2004A&A...427..873W} as well as some values from Table 5.2 of \cite{1989agna.book.....O}. We assume a maximum ratio of 1.5, determined by the ratio of the level statistical weights \citep{1989agna.book.....O} and use a corresponding minimum electron density of $10$\,cm$^{-3}$. The resulting variation in the collisonal line ratio as a function of logarithmic electron density is given in Figure \ref{neRats}. 
This electron density value is then used in the [O III] line ratio
\begin{equation}
		\frac{j_{\lambda 4959} + j_{\lambda 5007} }{j_{\lambda 4363}} = 							\frac{7.90\exp(3.29\times10^4/T)}{1+4.5\times10^{-4}n_{\rm{e}}/T^{1/2}}
\end{equation}
the [N II] line ratio
\begin{equation}
		\frac{j_{\lambda 6548} + j_{\lambda 6583} }{j_{\lambda 5755}} = 							\frac{8.23\exp(2.50\times10^4/T)}{1+4.4\times10^{-3}n_{\rm{e}}/T^{1/2}} 
\end{equation}
and the [Ne III] line ratio
\begin{equation}
		\frac{j_{\lambda 3869} + j_{\lambda 3968} }{j_{\lambda 3343}} = 							\frac{13.7\exp(4.30\times10^4/T)}{1+3.8\times10^{-5}n_{\rm{e}}/T^{1/2}} 
\end{equation}
low density limit ($n_{\rm{H}} <$ 10$^5$\,cm$^{-3}$) expressions to find the temperature \citep[e.g.][]{1989agna.book.....O}. 

Hence from one electron density sensitive line ratio and one temperature sensitive ratio an estimate of the electron density and temperature in the ionized gas can be obtained. With an estimate of the temperature and electron densities the pressure is calculated using equation \ref{pressureEqn}. This method has the advantage that the temperature of the ionized gas is determined directly, rather than being assumed to be 10000\,K as is the case for the radio diagnostics \citep[e.g.][]{2004A&A...414.1017T}.

\section{Testing}
The \textsc{torus} radiation hydrodynamics scheme is tested extensively in \cite{2012MNRAS.420..562H}. There are also numerous applications of the imaging and SED generating routines available in the literature \citep[e.g.][]{2004MNRAS.350..565H, 2004MNRAS.351.1134K}. Tests of the ratio diagnostics and the SED fitting are given here. 

\subsection{Diagnostic line ratio testing: the HII40 Lexington benchmark}
\label{lextestsec}
The HII40 Lexington benchmark involves, in one dimension and assuming no time evolution, calculating the temperature and ionization structure of the constant density gas surrounding a star at 40000\,K \citep{1995aelm.conf...83F}. The test includes no dust and the geometry of the cloud is a thick, constant density, spherical shell centered on the star with an inner radius 0.97\,pc ($3\times10^{18}$\,cm) from the star.  A list of the parameters used in the HII40 benchmark is given in Table \ref{LexingtonParams}.
Given the well studied nature of this temperature distribution \citep{1995aelm.conf...83F,2003MNRAS.340.1136E,2004MNRAS.348.1337W} it is an ideal test of the accuracy of our forbidden line ratio diagnostics.
We perform a photoionization calculation of a three dimensional version of the HII40 benchmark with the ionizing star set at the grid centre and apply forbidden line ratio diagnostics to synthetic images of the converged system. Given the large amount of neutral foreground material and the low dust fraction in the hot HII region, the dust to gas ratio in this test is set to a negligibly small value so that extinction does not have to be accounted for.
The model resolution, number of image pixels and slit size are chosen to match that of the RDI grids, images and emulated slit spectroscopy of BRCs used later in this paper. The parameters associated with synthetic imaging of this test are also given in Figure \ref{LexingtonParams}.
\begin{table}
 \centering
  \caption{Parameters for the HII40 Lexington benchmark which is used to test diagnostic line ratios.}
  \label{LexingtonParams}
  \begin{tabular}{@{}l c l@{}}
  \hline
   Variable (Unit) & Value & Description\\
 \hline
   $\textrm{T}_{\rm{eff}} \textrm{(K)}$ & 40000 & Source effective temperature \\
   ${R}_*(\rm{R}_\odot)$  & 18.67 & Source radius\\
   n$_{\rm{H}}$ (cm$^{-3}$) & 100 & Hydrogen number density\\
   log$_{10}$(He/H) & $-$1 & Helium abundance\\
   log$_{10}$(C/H) & $-$3.66 & Carbon abundance\\
   log$_{10}$(N/H) & $-$4.40 & Nitrogen abundance\\
   log$_{10}$(O/H) & $-$3.48 & Oxygen abundance\\
   log$_{10}$(Ne/H) & $-$4.30 & Neon abundance\\
   log$_{10}$(S/H) & $-$5.05 & Sulphur abundance\\
   L (pc$^3$) & $16^3$ & Computational domain volume\\
   $n_{\rm{cells}}$ & $128^3$ & Number of grid cells\\
   $n_{\rm{pix}} $ & $401^2$  & Synthetic image pixels\\
   $N_{\gamma}$ & $10^8$ & Photon packets used in \\
    & & synthetic image generation \\
   $L_{\rm{s}}$ (pixels) & 100 & Spectroscopic slit length\\
   $W_{\rm{s}}$ (pixels) & 2 & Spectroscopic slit width \\
\hline
\end{tabular}
\end{table}

\begin{figure}
		\hspace{-15pt}
		\includegraphics[width = 9cm]{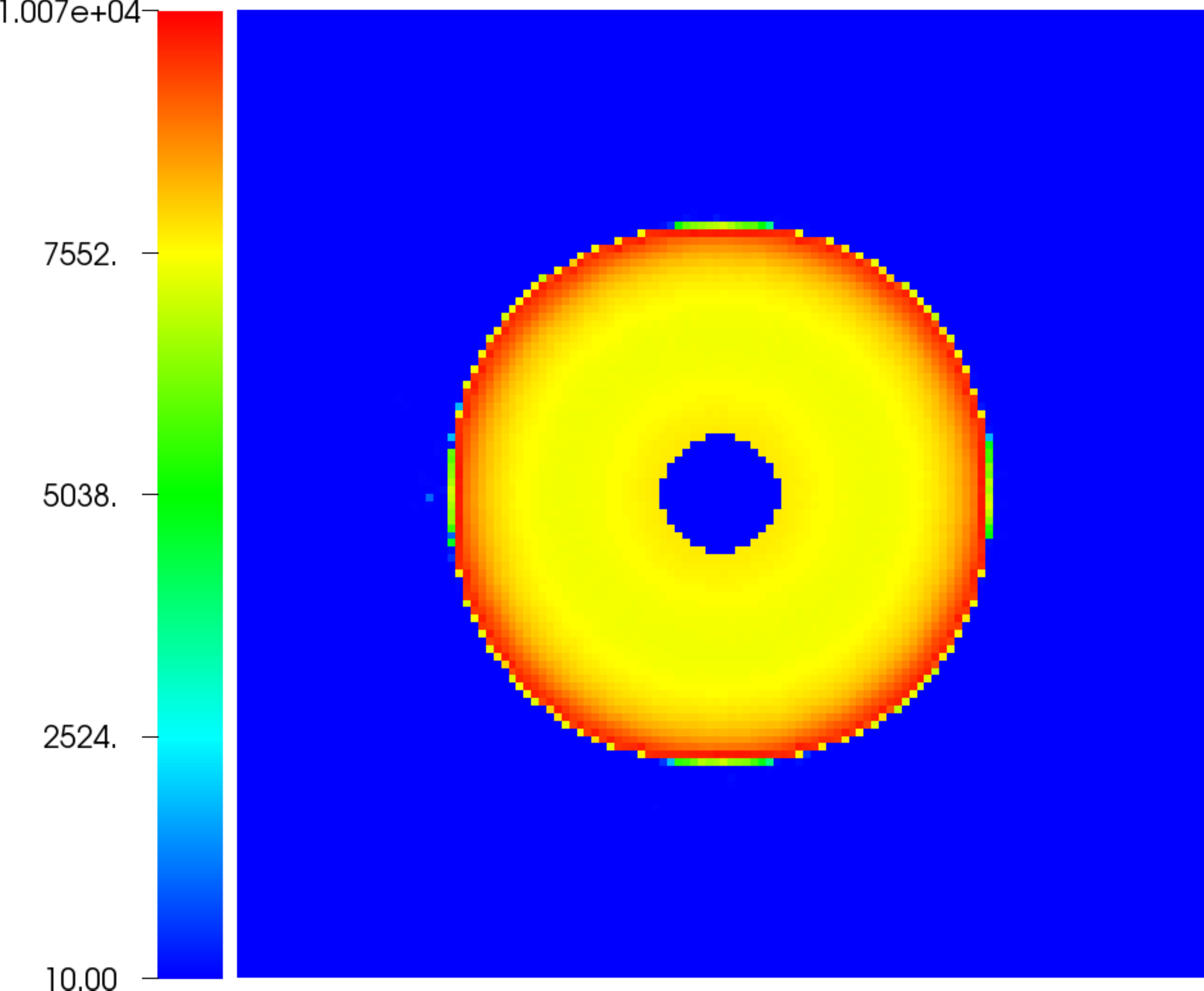}
		\caption{A slice through the three dimensional temperature distribution of the HII40 Lexington test. The grid is 16\,pc to a side.}
		\label{HII40temp}
\end{figure}

\begin{figure}
		\hspace{5pt}
		\includegraphics[width = 7.5cm]{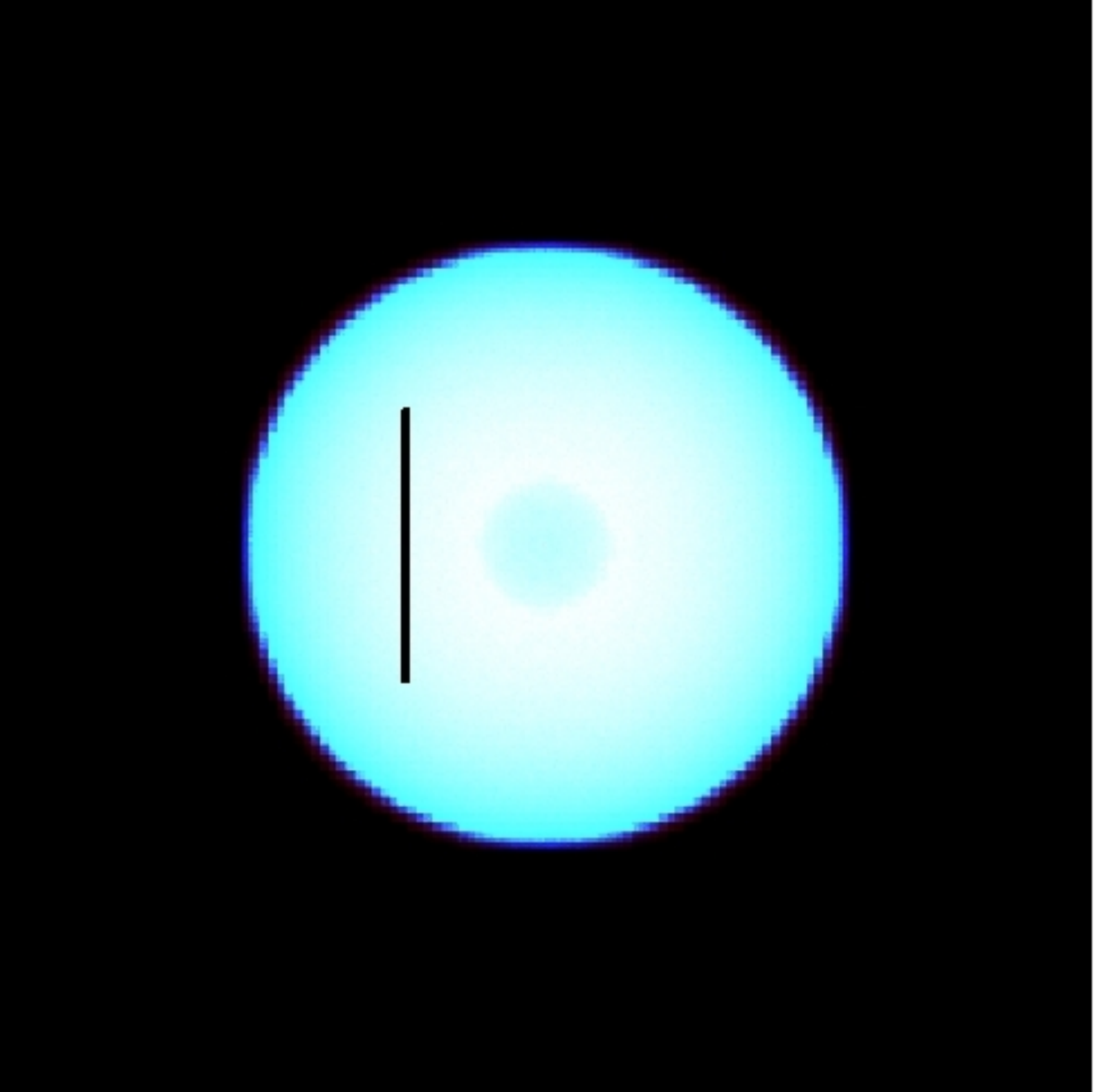}
		\caption{A colour composite image of the HII40 test using H$\alpha$ (red), the [O III] 5007\,\AA\, line (green) and the [Ne III] 3968\,\AA\, line (blue). Overlaid is the slit region used to emulate slit spectroscopy. The image side spans 16\,pc}
		\label{HII40img}
\end{figure}

The converged temperature state of the grid is given in Figure \ref{HII40temp} and a colour composite image comprising H$\alpha$ (red) the 5007\,\AA\,[O III]  line (green) and the 3968\,\AA\,[Ne III] line (blue) is given in Figure \ref{HII40img}, upon which is marked the region covered by the spectroscopic slit. In addition to the slit location in Figure \ref{HII40img}, we repeat the diagnostics for slit positions at $\pm1,2$ pixels in the $x$-direction to determine the level of uncertainty when using a single slit position to infer the conditions. The average electron density calculated using the [O II] diagnostic ratio and temperature calculated using the [O III], [N II] and [Ne III] diagnostic ratios are given in Table \ref{lextest}. 
The prescribed hydrogen number density is 100\,cm$^{-3}$, which is slightly higher than the inferred electron density. The HI fraction throughout the grid is not uniformly zero, however this accounts for less than 1\,cm$^{-3}$. Comparing the linear interpolation between our [O II] data points with a more sophisticated fit also only leads to an improvement of 3\,cm$^{-3}$. The remaining discrepancy is therefore believed to be characteristic of the diagnostic ratio.

Although there is some variation between the calculated temperatures, they all lie within the 7000-10000\,K range expected from Figure \ref{HII40temp}. The actual average HII region temperature in the region covered by the slit is 7769\,K, within 600\,K of each diagnostic and within 85\,K of the diagnostic average. Based on the uncertainties in individual diagnostics calculated in this test, further electron densities derived in this paper using the [O II] line ratio are conservatively calculated to the nearest 10\,cm$^{-3}$ and individual temperatures derived using the [O III], [N II] and [Ne III] line ratios are calculated to the nearest 10\,K. 

\begin{table}
\centering
  \caption{The Lexington HII40 test average conditions from diagnostics using a spectroscopic slit at 5 locations.}
  \label{lextest}
  \begin{tabular}{@{}l c c@{}}
  \hline
  Ratio & $n_{\rm{e}}$\,(cm$^{-3}$) & $T_{\rm{e}}$\,(K) \\
  \hline
  [O\,II],\,$j_{3729}/j_{3726}$ &  $88\pm2$ & $-$ \\
  & & 	\\
  $[\rm{O\, III}]$,\,($j_{5007} + j_{4959})/j_{4363}$ & $-$ & $7215\pm^6_7$ \\
  & & 	\\
  $[\rm{N\, II}]$,\,$(j_{6583} + j_{6548})/j_{5755}$ &  $-$ & $8366\pm^{4}_{3}$ \\
  & & 	\\
  $[\rm{Ne\, III}]$,\,$(j_{3968} + j_{3869})/j_{3343}$ &  $-$ & $7474\pm^6_9$ \\
\hline
\end{tabular}
\end{table}

\subsection{SED greybody fit testing: a cold uniform sphere}
\label{sedTest}
To test the cloud masses and temperatures derived by greybody SED fitting we consider the simple system that is a three dimensional uniform density sphere at constant temperature, surrounded by vacuum. We choose a density of 100 hydrogen atoms per cubic centimetre and a cloud radius of 1.6\,pc \citep[equivalent to the cut-off radius of the starting Bonnor-Ebert sphere in the models of][]{2012MNRAS.420..562H}. The surrounding vacuum is prescribed a number density of 10$^{-30}$ hydrogen atoms per cubic centimetre so that it will only negligibly contribute to the system SED. The model is prescribed a uniform temperature distribution of 10\,K throughout the grid.

\begin{figure}
		\hspace{-20pt}
		\includegraphics[width=9cm]{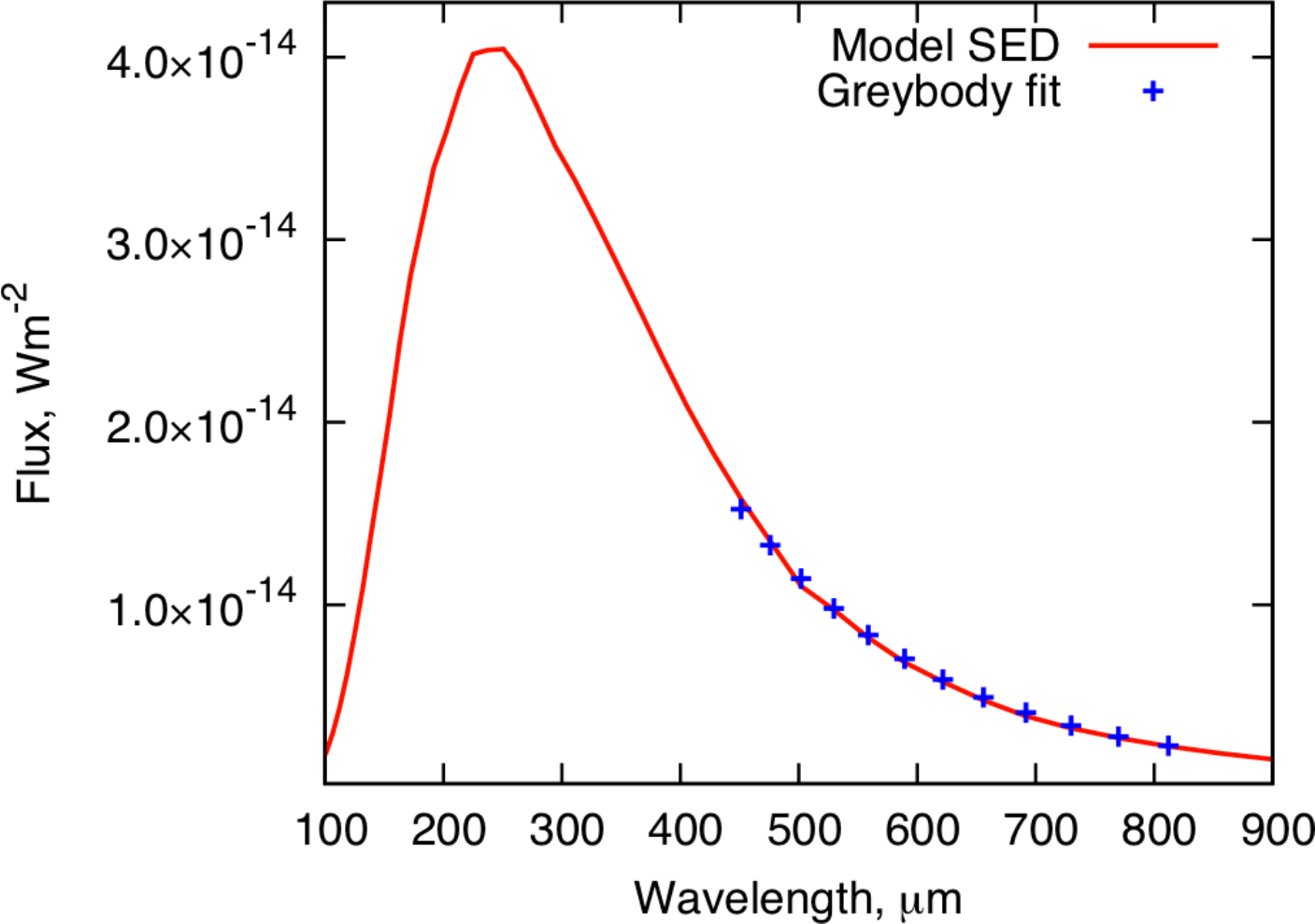}
		\caption{The uniform cloud model SED (red line) and the greybody fit (blue crosses).}
		\label{SEDfit}
\end{figure}

We produce an SED for an observer situated 1000\,pc from the grid. The spectrum is fitted in the manner described in section \ref{greyFitting} from 450\,$\mu$m to 850\,$\mu$m following \cite{2004A&A...414.1017T} and is shown in Figure \ref{SEDfit}. The resulting calculated temperature is 9.5\,K, within 5\% of the prescribed value.
The total mass of this cloud is 42.4\,M$_{\odot}$. We use a range of values for C$_{\nu}$ to calculate a fitted mass from the SED, the results of which are given in Table \ref{isospheremasses}. We find that the most appropriate value of C$_{\nu}$ for number densities of order 100\,cm$^{-3}$ at low temperatures is 214\,g\,cm$^{-2}$. The average column density, calculated by integrating the column density over the uniform sphere and dividing by the projected area,  is $1.3\times10^{20}$\,cm$^{-2}$, the optical depth calculated using equation \ref{getTau} is $2.2\times10^{-6}$ and the corresponding $C_{\nu}$ value using equation \ref{cnu} is 131\,g\,cm$^{-2}$, just under a factor of 2 smaller than our inferred value. The reason for the discrepancy between the theoretical and inferred values is likely due to averaging the large variations in column density across the sphere, from its maximum value at the centre to approximately zero at the edge. Given that densities of order 100\,$m_{\rm{H}}$\,cm$^{-3}$ are typical of the neutral clouds in the RDI models, we retain the presently inferred value of $C_{\nu}=214$\,g\,cm$^{-2}$ in further SED fitting in this paper. 

\begin{table}
\centering
  \caption{Cold uniform sphere fitted masses.}
  \label{isospheremasses}
  \begin{tabular}{@{}l c@{}}
  \hline
   C$_{\nu}$ (g\,cm$^{-2}$) & Mass M$_{\odot}$\\
 \hline
  50 & 9.9\\
  100 & 19.8\\
  200 & 39.6\\
  214 & 42.4\\
  250 & 49.5\\
\hline
\end{tabular}
\end{table}

\section{Results and Discussion}

\subsection{The numerical models}
\label{tempGrids}
We add atomic species to the final state (after 200\,kyr of evolution) of the RDI models in \cite{2012MNRAS.420..562H} and perform photoionization calculations. The three models, each of which comprised a Bonnor-Ebert sphere (BES) exposed to a different level of ionizing flux (high, medium and low), underwent thermal and photo-evaporative compression.
The low flux model exhibited thermal compression of the BES and intermediate strength photo-evaporative flow, resulting in a type A bow. Instabilities in the ionization front also led to the formation of finger like objects in the wings of the low flux model. The medium flux model quickly accumulated a dense shell of material resulting in a strong photo-evaporative flow that compressed the cloud to a type B-C cometary bow. The high flux model rapidly established pressure balance and exhibited only weak photo-evaporative flow, leaving what resembles a type B bow. A summary of the key parameters of these models is given in Table \ref{RDIparams}.

\begin{figure}
	\hspace{-16.5pt}
	\includegraphics[width=8.84cm]{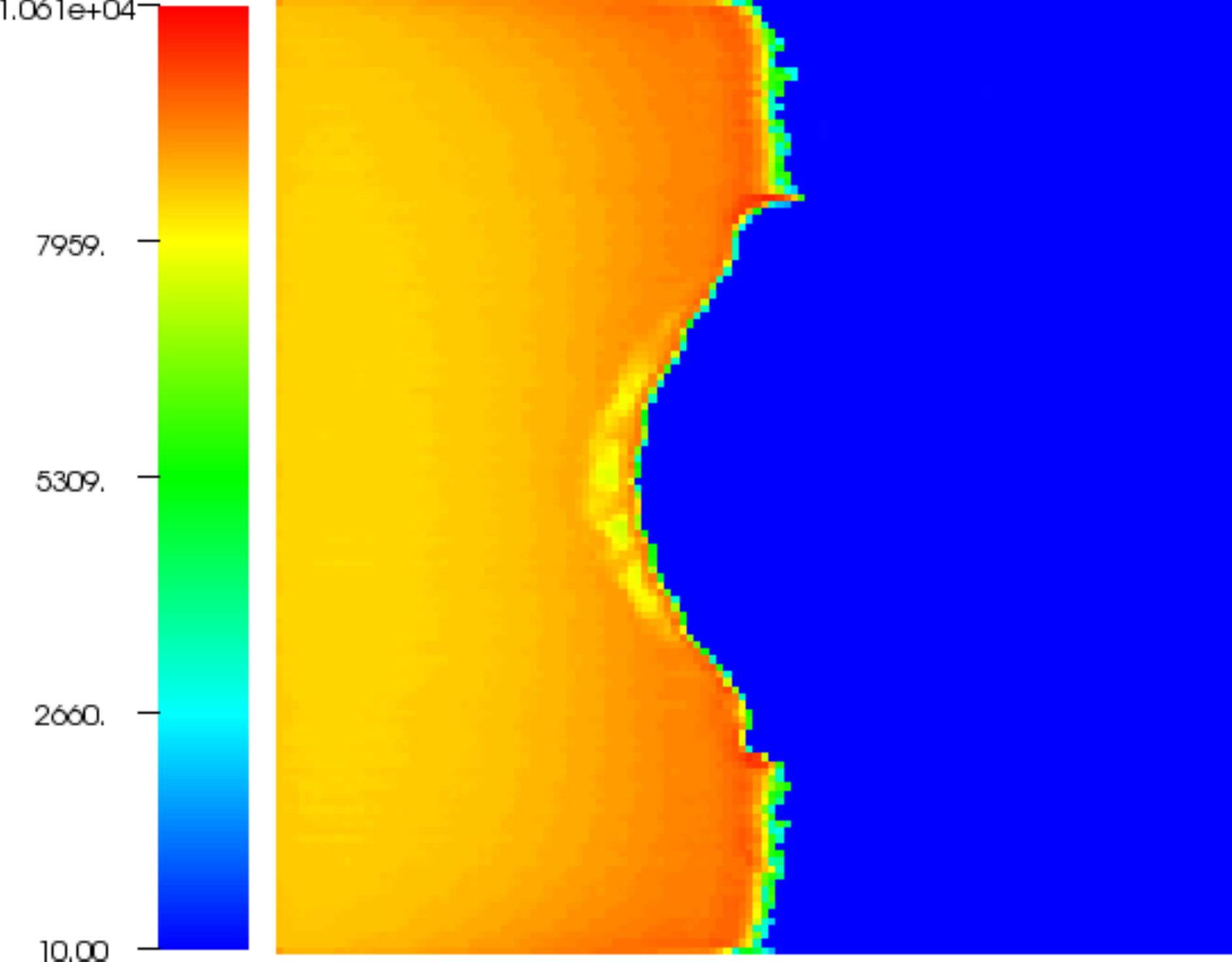}

	\vspace{5pt}
	\hspace{40pt}
	\includegraphics[width=6.85cm]{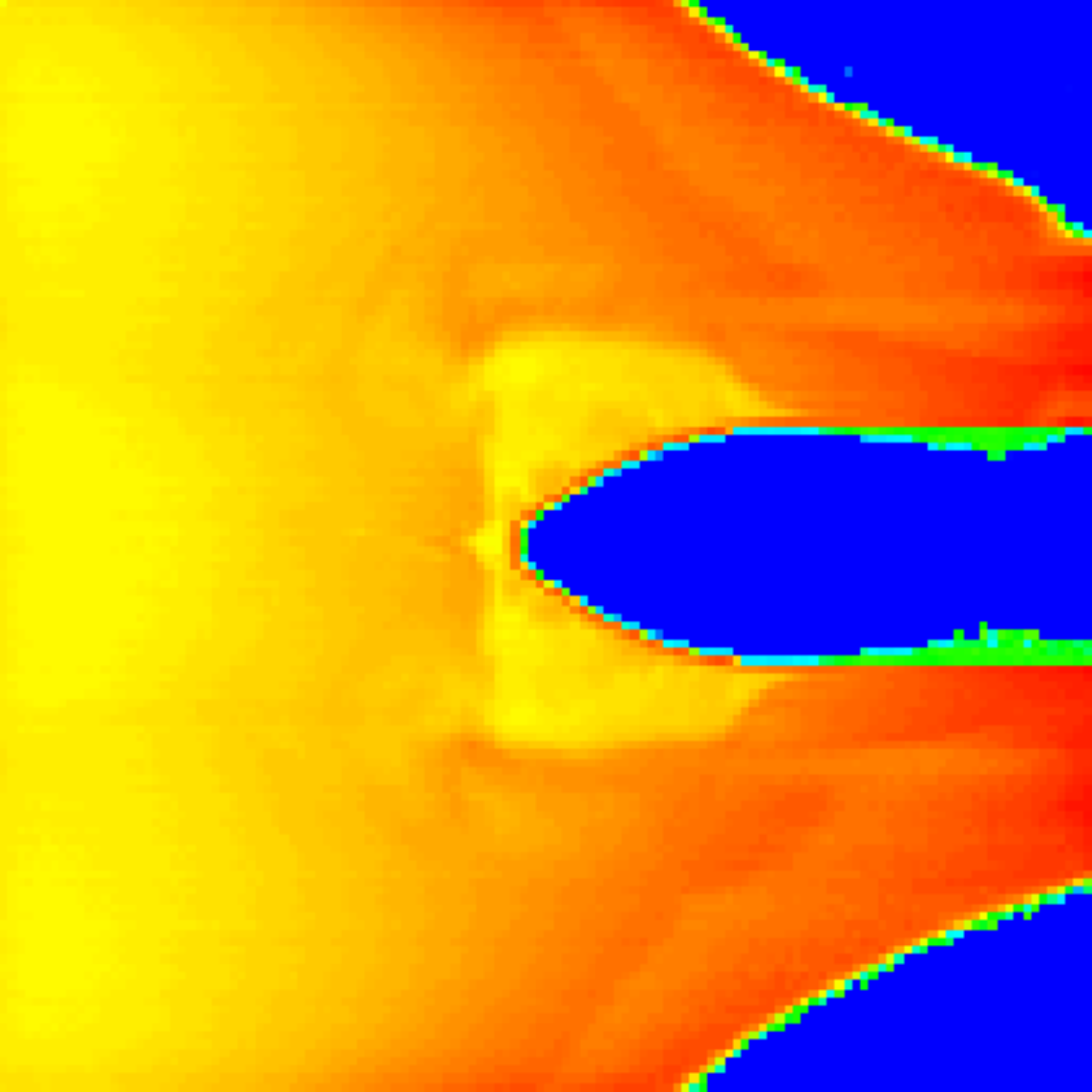}

	\vspace{5pt}
	\hspace{40pt}
	\includegraphics[width=6.85cm]{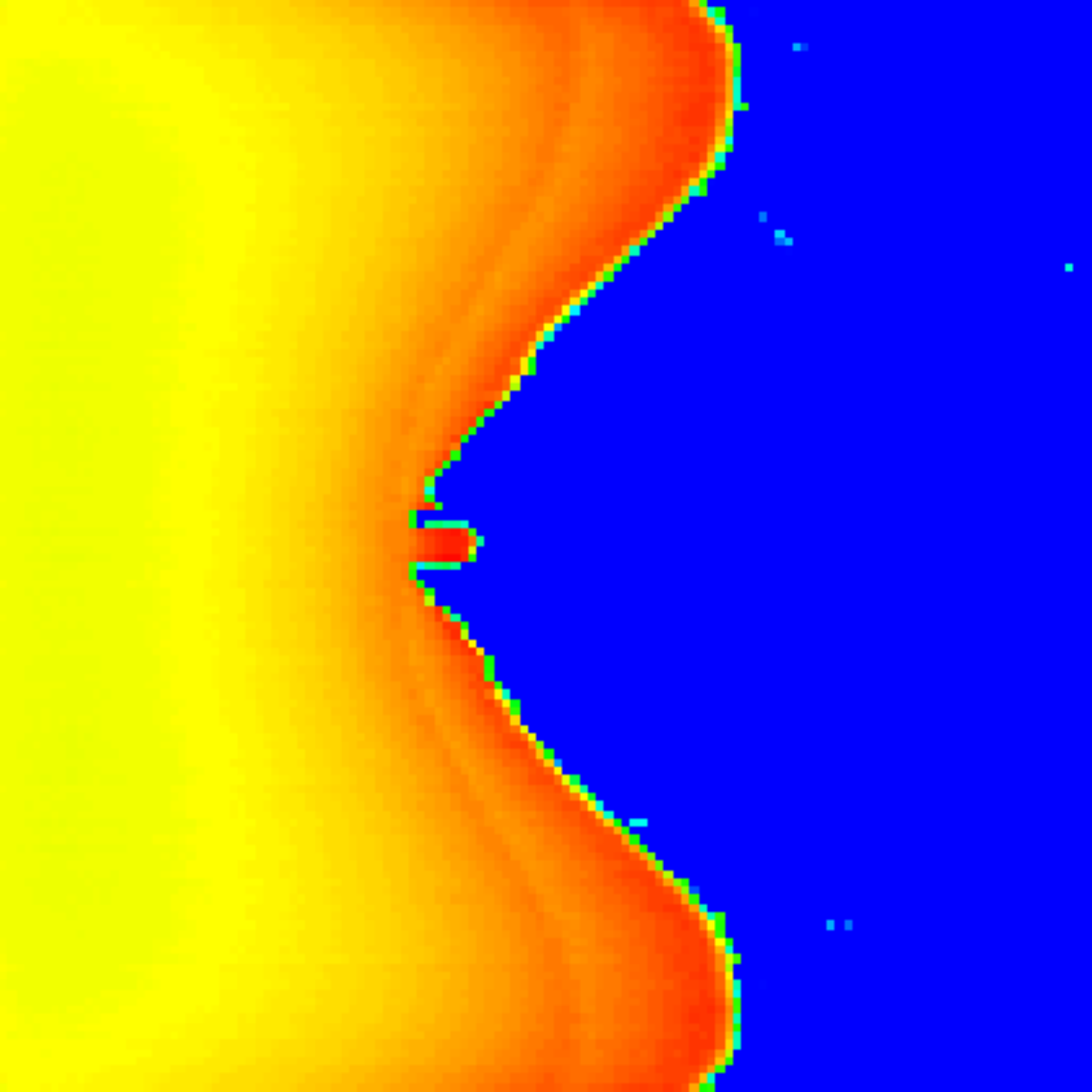}
	\caption{The temperature distribution (in Kelvin) for each model, low flux to high flux from top to bottom. The grid is 4.87\,pc to a side.}
	\label{temperatures}
\end{figure}

The temperature state of each model following the photoionization calculation with additional atomic chemistry is given in Figure \ref{temperatures}. In each model, typical ionized gas temperatures range from $7000-10000$\,K and neutral temperatures are essentially uniform at 10\,K. Of the cells with an HI ionization fraction greater than 0.5 and temperature less than 1500\,K: 97\%, 96\% and 98\% are at 10\,K for the low, medium and high flux models respectively. By mass 96\%, 95\% and 98\% of the cloud is at 10\,K for the low, medium and high flux models respectively. Hot spots in the neutral gas of the bottom frame of Figure \ref{temperatures} are artifacts due to Monte Carlo sampling where, on rare occasion, high energy photon packets propagate a significant distance into the neutral gas. However, these spots are too hot for dust to survive and are away from the region over which we analyze the IBL so will not affect our synthetic SEDs or long-slit spectroscopy.

The low and medium flux models have retained a cold neutral gas morphology similar to that at the end of their radiation hydrodynamic evolution. 
Because the disruption to the original BES significantly modified the density distribution in the low and medium flux regimes by excavating and accumulating material, the addition of metals (and therefore forbidden line cooling) and more sophisticated thermal balance have little effect on the extent of the ionized region. This is not the case for the high flux model, which weakly modified the density distribution, achieving early pressure balance and establishing only weak photo-evaporative flow. As a result the enhanced cooling has shifted the ionization front away from the cloud structure that was formed in the radiation hydrodynamics calculation to a new position similar to that of the low flux model. This additional cooling in weakly disrupted gas is also responsible for the  cool regions towards the edge of the medium flux grid.
These alterations to the extent of the ionized regions hint at the importance of including additional atomic processes such as forbidden line cooling in radiation hydrodynamics calculations. Although the effect is small in the low and medium flux model, any diagnostic procedure applied to the high flux model will not be examining the pressure balanced system generated in \cite{2012MNRAS.420..562H}, but a new one in which D-type expansion about the HII region is yet to occur.

The low and medium flux models both exhibit slightly cooler regions where strong photo-evaporative flow occurs. This is the yellow region mirroring the central part of the bow of the top frame of Figure \ref{temperatures} for the low flux model and the shovel shaped yellow region about the tip of the cometary object in the middle frame of Figure \ref{temperatures} for the medium flux model.  

\begin{table*}
 \centering
 %\begin{minipage}{180mm}
  \caption{Key RDI model parameters.}
  \label{RDIparams}
  \begin{tabular}{@{}l c l@{}}
  \hline
   Variable (Unit) & Value & Description\\
  \hline
 $D_{\rm{low}}$ (pc) & $(-10.68,0,0)$ & Source position (low flux) \\
 $D_{\rm{med}}$ (pc) & $(-4.78,0,0)$ & Source position (medium flux) \\ 
 $D_{\rm{hi}}$ (pc) & $(-3.38,0,0)$ & Source position (high flux)\\
 $T_{*}$ (K) & 40000 & Source effective temperature \\
 $R_{*}$ (R$_{\odot}$) & 10 & Source radius \\
 $\Phi_{\rm{low}}$ (cm$^{-2}$\,s$^{-1}$) & $9\times10^8$ & Low ionizing flux at the left grid edge \\
 $\Phi_{\rm{med}}$ (cm$^{-2}$\,s$^{-1}$) & $4.5\times10^9$ & Medium ionizing flux at the left grid edge \\
 $\Phi_{\rm{hi}}$ (cm$^{-2}$\,s$^{-1}$) & $9\times10^9$ & High ionizing flux at the left grid edge \\
 L(pc$^3$) & $4.87^3$ & Grid size\\
 $n_{\rm{cells}}$ & $128^3$ & Number of grid cells\\
 $n_{\rm{pix}} $ & 401$^2$ & Synthetic image pixels\\
 $\theta_{\rm{pix}} ($\arcsec$)$ & 2.5 & Angular width per pixel \\
 $N_{\gamma}$ & $10^8$ & Photons packets used in synthetic image\\
 $D_{\rm{o}}$ (pc)& 1000 & Observer distance \\
 dust/gas  & $1\times10^{-2}$ &Dust to gas ratio \\
 log$_{10}$(He/H) & $-$1 & Helium abundance\\
 log$_{10}$(C/H) & $-$3.66 & Carbon abundance\\
 log$_{10}$(N/H) & $-$4.40 & Nitrogen abundance\\
 log$_{10}$(O/H) & $-$3.48 & Oxygen abundance\\
 log$_{10}$(Ne/H) & $-$4.30 & Neon abundance\\
 log$_{10}$(S/H) & $-$5.05 & Sulphur abundance\\
  \hline
\end{tabular}
%\end{minipage}
\end{table*}

\subsection{Optical Image Morphology}
\begin{figure}
	\includegraphics[width=7.1cm]{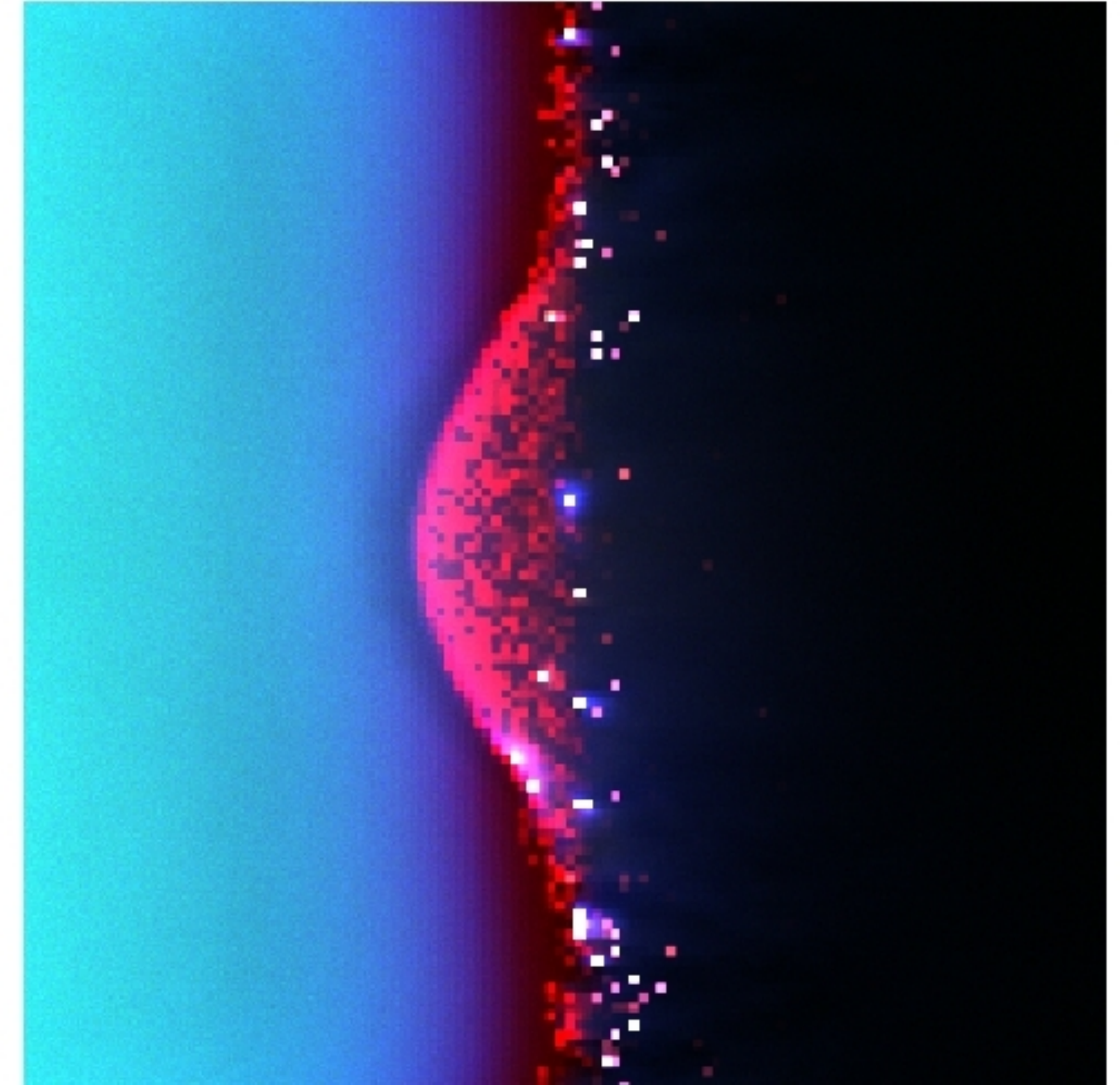}

	\vspace{5pt}
	\includegraphics[width=7.22cm]{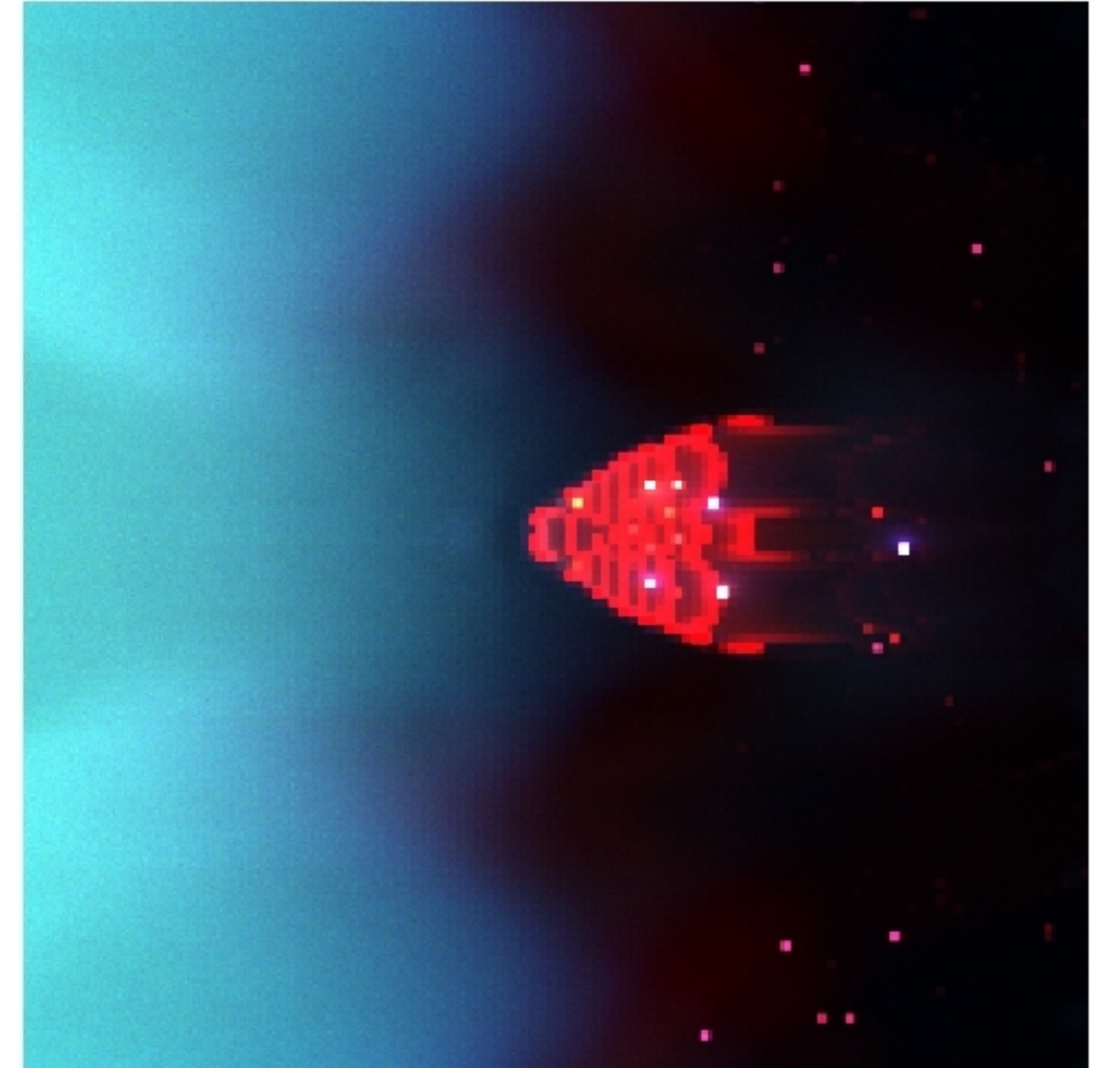}

	\vspace{5pt}
	\includegraphics[width=7.1cm]{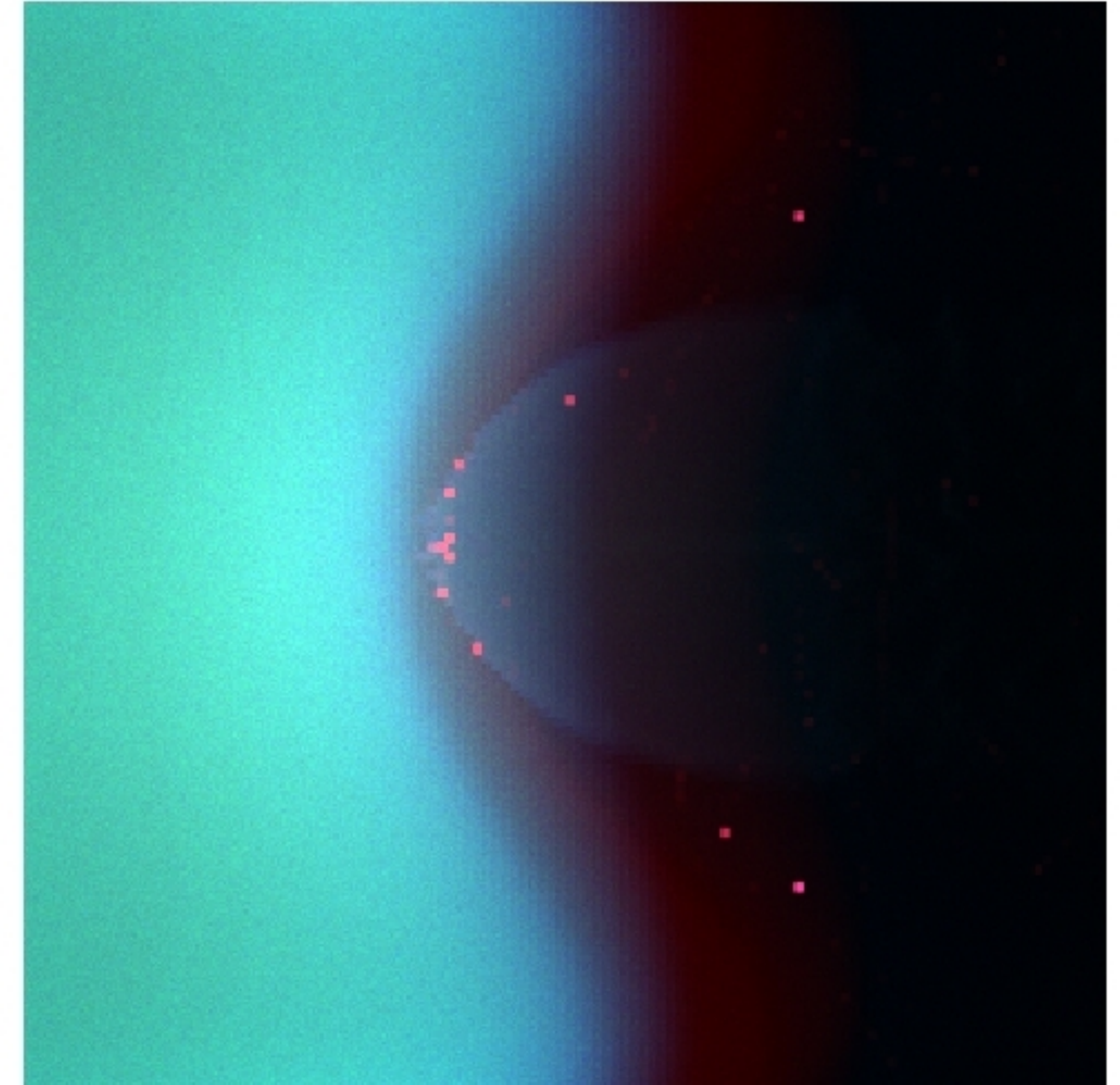}
	\caption{Composite colour images of the low, medium and high flux models from top to bottom using H$\alpha$ (red), the 5007\,\AA\, [O III] line (green),  and the 3968\,\AA\, [Ne III] line (blue). Each image side spans 4.87\,pc.}
	\label{Imgs}
\end{figure}

\begin{figure}
	\hspace{8pt}
	\includegraphics[width=7cm]{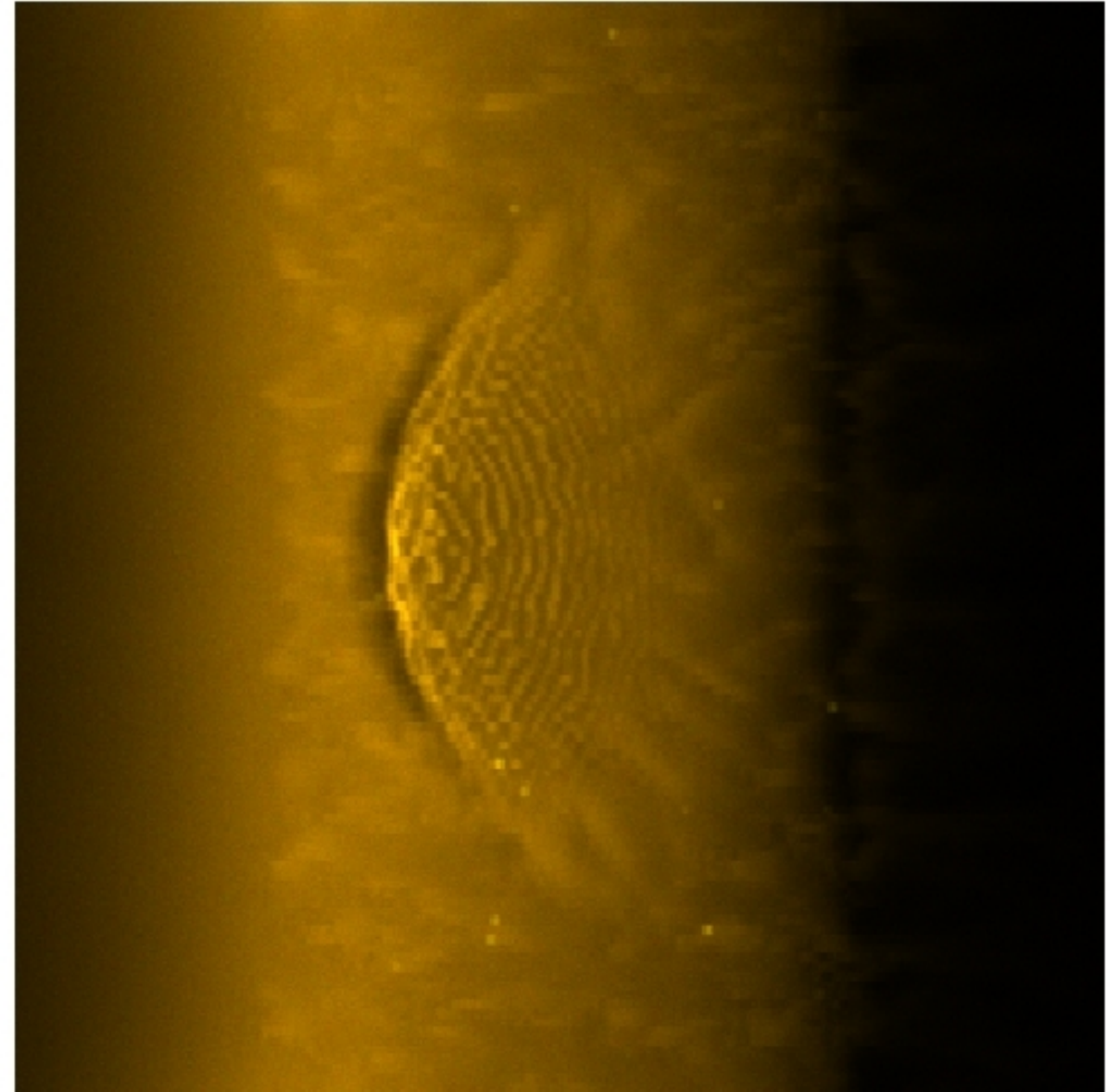}
	\caption{A two colour image of the low flux model for an observer inclined thirty degrees relative to the planar ionization front. This image is constructed using the electron density sensitive 3729\,\AA\, (red) and 3726\,\AA\ (green), [O II] lines. The image side spans 4.87\,pc.}
	\label{lowInclined}
\end{figure}

A synthetic image of the low flux model using H$\alpha$ (red), the 5007\,\AA\,[O III] line (green) and the 3968\,\AA\, [Ne III] line (blue) is given in the top frame of Figure \ref{Imgs}. 
The object clearly resembles a class A, moderately curved, bright rimmed cloud. H$\alpha$ dominates the emission along the rim, making it appear red. The signature of the photo-evaporative flow is the slightly darker region opposite the bright rim, which has been excavated to slightly lower densities due to the motion of the flow. The hot (blue) spots are due to Monte Carlo sampling as explained in section \ref{tempGrids}.
From this edge-on inclination the fingers due to instabilities (see section \ref{tempGrids}) are not clear because along a given line of sight there are multiple trunks and the average column density is approximately constant. Furthermore, the enhanced cooling due to forbidden line processes means that the ionized regions between the trunks are now neutral, suppressing emission from them. A two colour image in the electron density sensitive lines of [O II] where the observer is inclined by 30 degrees relative to the ionization front is shown in Figure \ref{lowInclined}.  Here the fingers and dark photo-evaporative flow region are much more clearly illustrated.  The morphological resemblance between this image and that in Figure 1 of \cite{2011PASJ...63..795C} is notable, despite the fact that the additional cooling has made the fingers harder to discern. In \cite{2011PASJ...63..795C} it was suggested that the finger like objects in the wings of the BRC were due to instability, this image and the radiation hydrodynamic model it is generated from qualitatively support this hypothesis. 

\begin{figure}
	\hspace{8pt}
	\includegraphics[width=7cm]{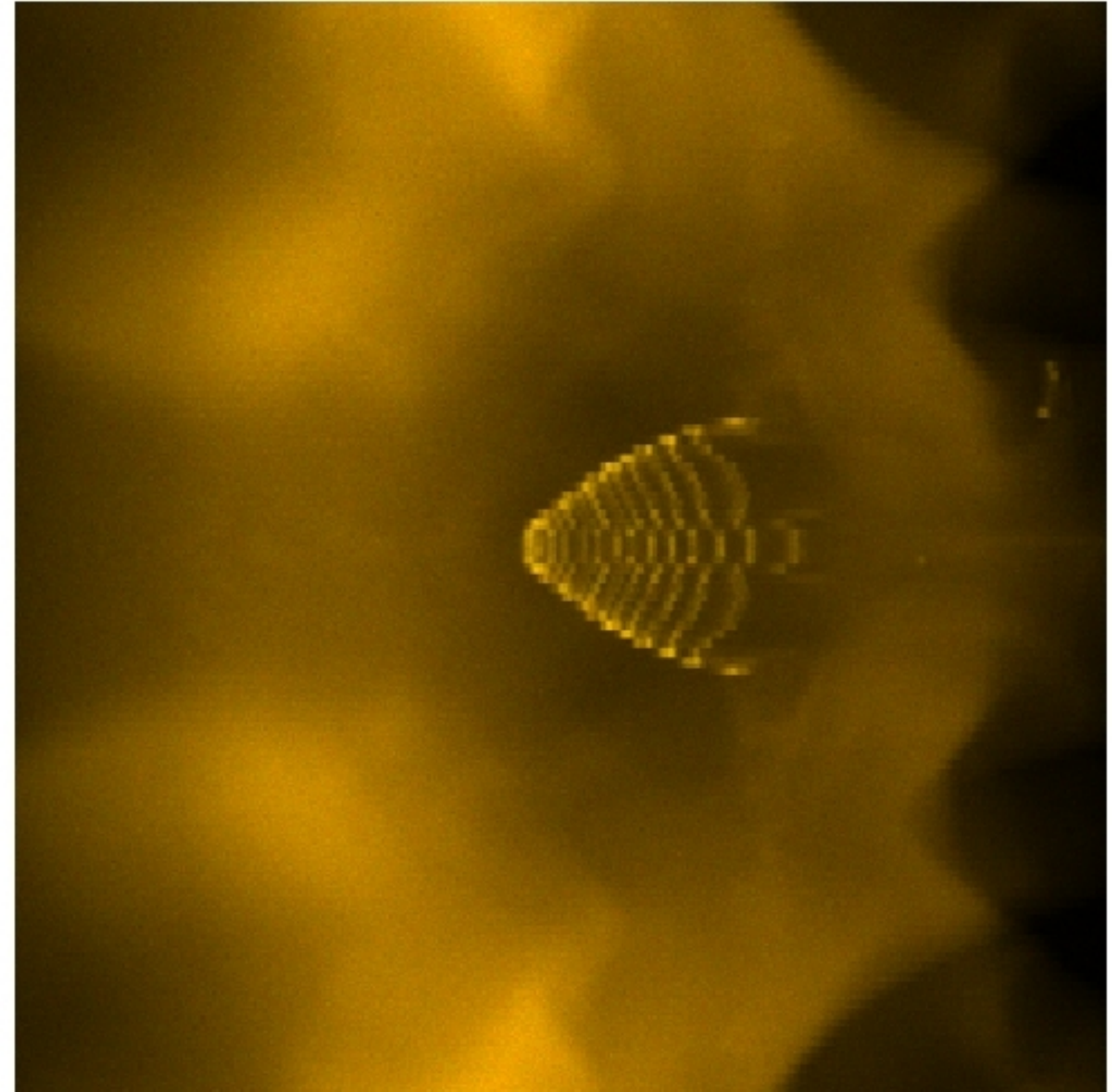}
	\caption{A two colour image of the medium flux model for an observer inclined thirty degrees relative to the planar ionization front. This image is constructed using the electron density sensitive 3729\,\AA\, (red) and 3726\,\AA\, (green) [O II] lines. The image side spans 4.87\,pc}
	\label{medInclined}
\end{figure}

A synthetic image of the medium flux model using H$\alpha$ (red), the 5007\,\AA\,[O III] line (green) and the 3968\,\AA\, [Ne III] line (blue) is given in the middle frame of Figure \ref{Imgs}. This object resembles a type B-C BRC, having an almost cometary appearance that is again dominated by H$\alpha$ emission. The ribbing effect interior to the cometary structure is a result of the low resolution used in the radiation hydrodynamics calculation and has no physical significance. The bright ribs are the cells exposed to the direct stellar radiation field and the darker regions are shielded cells, for which radiative heating comes primarily from the diffuse field. Despite being known to have a stronger photo-evaporative flow than the low flux model, it is difficult to distinguish between a dark excavated region and neutral foreground material. Again, we take an image at thirty degrees inclined relative to the ionization front to illustrate this point in Figure \ref{medInclined}, in which the dark foreground material starts moving out of view. This image suggests that the BRC is situated in the bottom of a basin of ionized gas. Furthermore, a darker region excavated by the photo-evaporative flow is visible around the BRC in these electron density sensitive diagnostic lines.

A synthetic image of the high flux model using H$\alpha$ (red), the 5007\,\AA\,[O III] line (green) and the 3968\,\AA\, [Ne III] line (blue) is given in the bottom frame of Figure \ref{Imgs}. Note that the radiation hydrodynamics model did not significantly alter the starting density distribution for this model, accumulating only a small amount of material before achieving pressure balance and establishing a weak photo-evaporative flow. As a result the enhanced cooling due to forbidden line radiation and the more complex thermal balance calculation have increased the size of the neutral gas region, i.e. the Str\"{o}mgren radius is smaller. The resulting imaged object has the curvature of a class A BRC, though it is not actually bright rimmed, beneath which hints of the old neutral cloud structure can be seen.

\subsection{Neutral gas properties}
\label{greybodyAnalysis}
\begin{figure}
	\hspace{-20pt}
	\includegraphics[width=9.2cm]{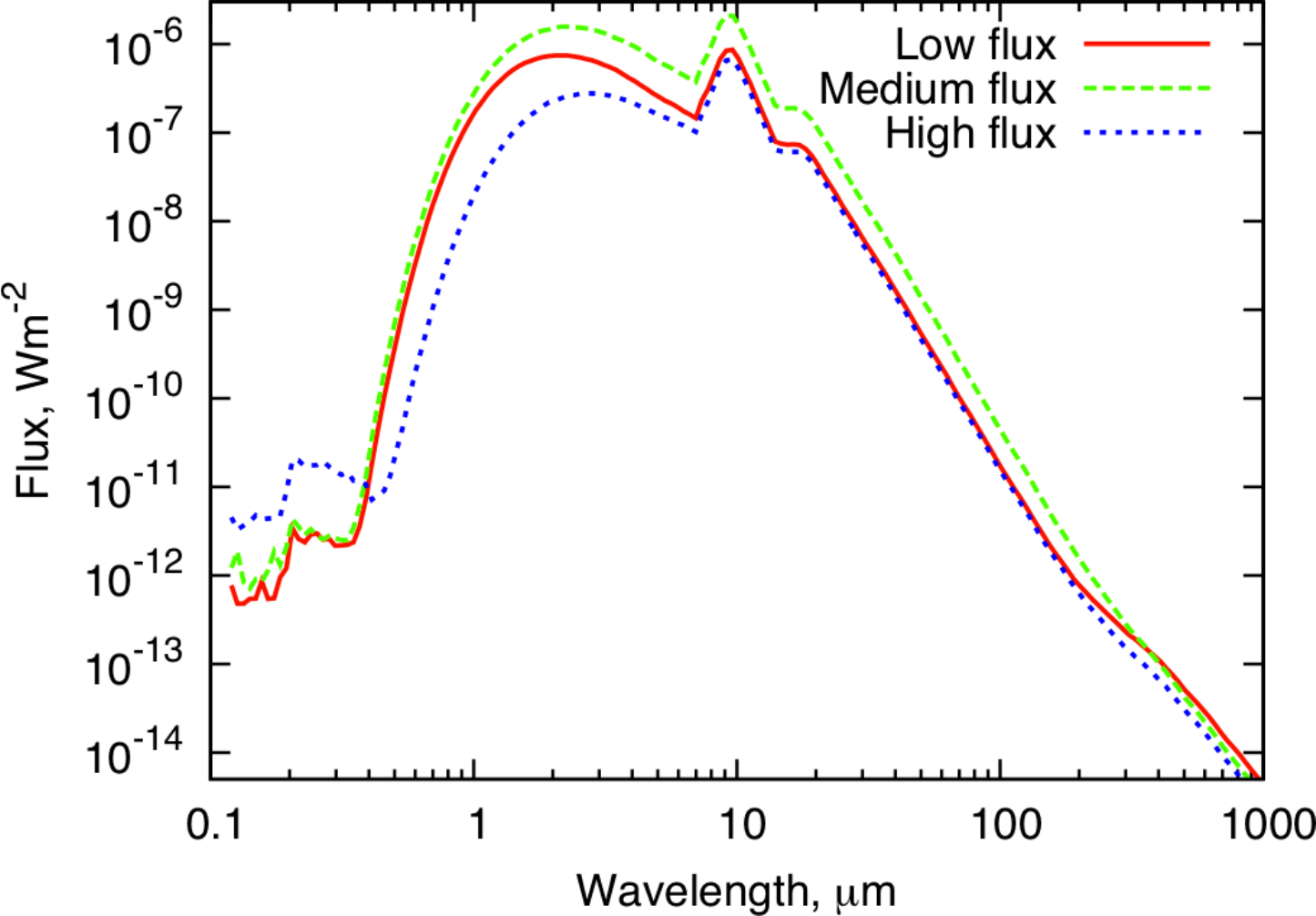}
	\caption{A a comparison of the low, medium and high flux model spectra.}
	\label{fittedSpectra}
\end{figure}

Spectral energy distributions which include dust and free-free emissivities were generated for each converged grid, these are shown in Figure \ref{fittedSpectra}. Short of the near UV regime (from approximately 380\,nm bluewards) scattered stellar photons dominate the signal and the flux decreases in accordance with the star's distance from the grid, i.e. the strongest signal is from the high flux model and the weakest is from the low flux model. Over most of the rest of the spectrum, where direct thermal radiation dominates the SED, the medium flux model consistently has the strongest signal, followed by the low then high flux models. At wavelengths greater than about 370\,$\mu$m the low flux spectrum has the strongest signal due to the greater spatial extent of the cloud and therefore more widespread dust emission.
The 10\,$\mu$m silicate feature is clearly visible in emission in all SEDs.

The dust temperatures and cloud masses are derived using the greybody fitting method described in section \ref{greyFitting}. We use a value of 214\,g\,cm$^{-2}$ for $C_\nu$ and fit the spectrum between 450 and 850\,$\mu$m following \cite{2004A&A...414.1017T}. 
The resulting temperatures and cloud masses for each system are given in Table \ref{SEDfitted}. 
The temperatures are all close to the dominant (that for at least 95\% by mass) neutral cloud temperature of 10\,K. The derived temperature is increased slightly by $1-2$\,K by warmer dust at the interior edge of the bright rim. 

\begin{table*}
\centering
  \caption{Neutral cloud properties from SED fitting. Included are two choices of $C_{\nu}$ and the masses calculated using the dominant temperature of 10\,K.}
  \label{SEDfitted}
  \begin{tabular}{@{}l c c c c c@{}}
  \hline
   Model & Dust & Cloud mass, ($\rm{M}_{\odot}$) & Cloud mass, ($\rm{M}_{\odot}$) & Cloud mass, ($\rm{M}_{\odot}$) & Cloud mass  \\
   & temperature\,(K)  &$C_{\nu}$ = 214\,g\,cm$^{-2}$& $C_{\nu}$ = 50\,g\,cm$^{-2}$ & $C_{\nu}$ = 214/50\,g\,cm$^{-2}$, T = 10\,K & from grid\,($\rm{M}_{\odot}$) \\
 \hline
 Low flux & 11 &  $144$ & 34 & 159/37 & 74\\
 Medium flux & 12 & $75$ & 18 & 115/27 & 21\\
 High flux & 11 & $76$ &  18 & 90/21 & 77\\
\hline
\end{tabular}
\end{table*}

The calculated neutral mass in the low and medium flux cases are overestimated by a factor of approximately 2 and 3.6 respectively. This is due to the standard practice of using a fixed value of $C_{\nu}$ \citep[e.g.][]{2004A&A...414.1017T,2008A&A...477..557M} which is probably too large in this case. The correct mass would have been inferred using values of 110 and 60\,g\,cm$^{-2}$ for $C_{\nu}$ in the low and medium flux mass calculations respectively. 
The inferred mass of the high flux cloud is close to the known value on the computational grid. This is because lower densities were attained in this model making the adopted value of $C_{\nu}$ more appropriate. 

Cloud masses calculated using the commonly adopted value of 50\,g\,cm$^{-2}$ for $C_{\nu}$ \citep[e.g.][]{2004A&A...414.1017T} are also given in Table \ref{SEDfitted}. These give a reasonable estimate of the medium flux mass but underestimate the low flux mass by over a factor of two. The large error in the high flux mass when using the lower value of $C_{\nu}$ is less significant due to the differences discussed in section \ref{tempGrids}.

The sensitivity of this calculation to the dust temperature at low values is also illustrated in Table \ref{SEDfitted}, where the calculated mass assuming a cloud temperature of 10\,K (the dominant temperature in the neutral cloud, see section \ref{tempGrids}) is also given for both values of $C_{\nu}$ used here. In the low and high flux cases the 1\,K variation in temperature modifies the calculated masses by a factor in the range $8-16$\,\%. The 2\,K variation in the medium flux case leads to a change in the calculated mass by about 35\,\%.

This mass calculation is already treated with caution \citep[e.g.][]{2001ApJ...552..601K,2004A&A...414.1017T,2008A&A...477..557M} these results suggest that using a fixed value of $C_{\nu}$ for a range of clouds of different class will induce errors of a factor up to around 3.6 in some of the cloud masses and that the calculation is sensitive to temperature at values around 10\,K with a typical difference in the mass of about 15\,\%\,K$^{-1}$.

\subsection{Radio analysis}
\label{radioAnalysis}
20\,cm radio images are used to determine the photoionizing flux, IBL electron density and mass loss rate following the discussion in section \ref{radio}. Synthetic images at 20\,cm are generated, smoothed and subjected to noise in the manner described in sections \ref{imgseds} and \ref{radImg} to be representative of the resolution and noise level of 30 second exposures of the VLA type B, C and D configurations. 30 second exposures using the D configuration are typical of those used in the NRAO VLA Sky Survey (NVSS) \citep{1998AJ....115.1693C}. We assume a bandwidth of 43\,MHz and the use of 26 antennae. The rms noise associated with these observations is 0.524\,mJy\,beam$^{-1}$. Table \ref{VLAstats} summarises the HPBW value and rms pixel noise for each configuration.
\begin{table}
\centering
  \caption{HPBW sizes and pixel noise levels for the images based on 30\,s VLA exposures.}
  \label{VLAstats}
  \begin{tabular}{@{}l c c@{}}
  \hline
   Configuration & HPBW & rms pixel noise (mJy\,pixel$^{-1}$)\\
 \hline
 B & 3.9$\arcsec$  &0.27\\
 C & 12.5$\arcsec$  &$2.7\times10^{-2}$\\
 D & 44$\arcsec$ &$2.2\times10^{-3}$\\
\hline
\end{tabular}
\end{table}
The regions used to derive the integrated IBL flux are the circular regions labelled `B' on H$\alpha$ images in Figure \ref{ha_slits}. These all have an angular diameter of $1\arcmin$. The cloud radii are those of the circular regions marked `A' on Figure \ref{ha_slits} and are 1.5, 0.6 and 1.3\,pc for the low, medium and high flux models respectively. 

\begin{figure*}
	\includegraphics[width=4.2cm]{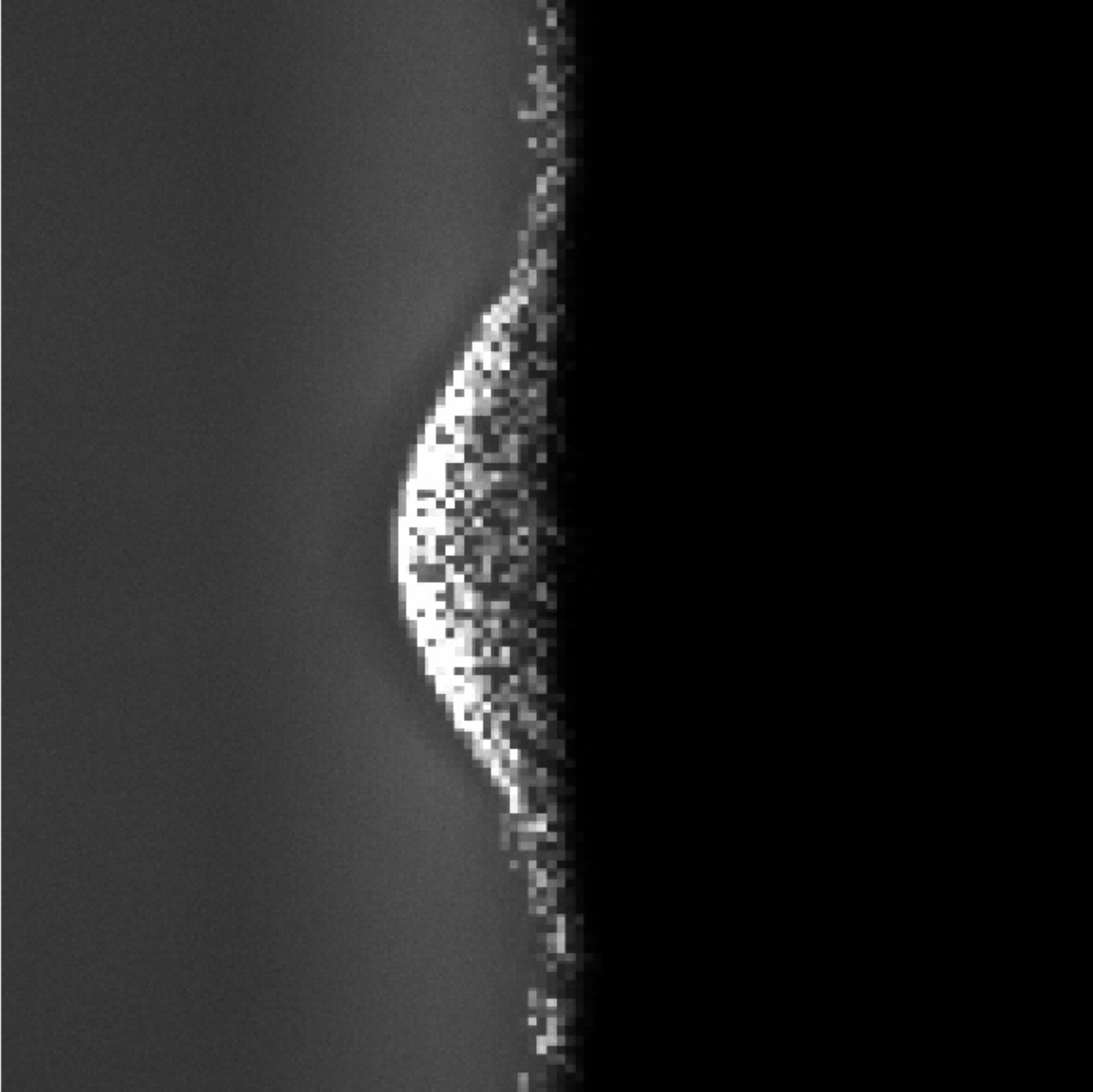}
	\includegraphics[width=4.2cm]{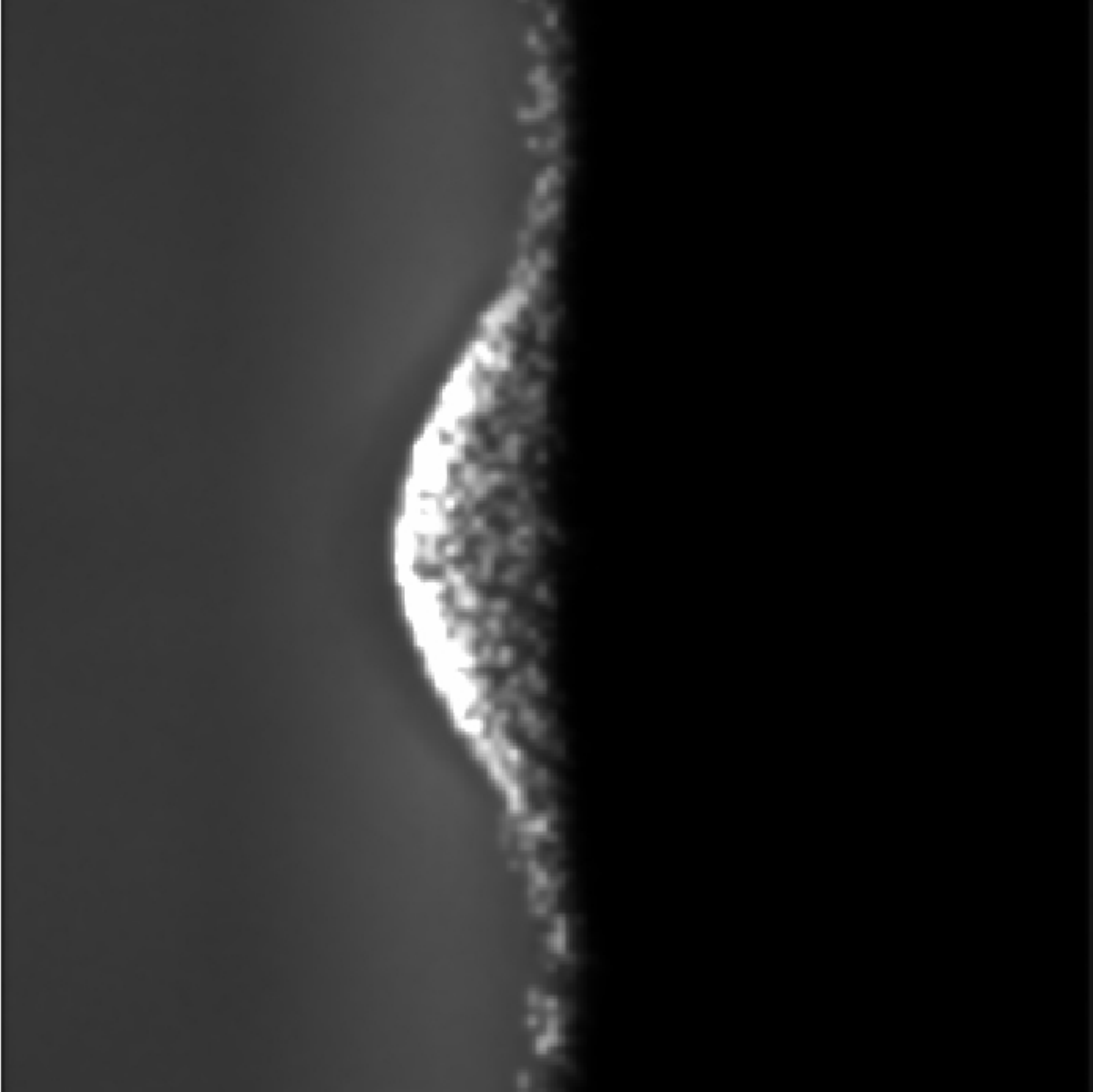}
	\includegraphics[width=4.2cm]{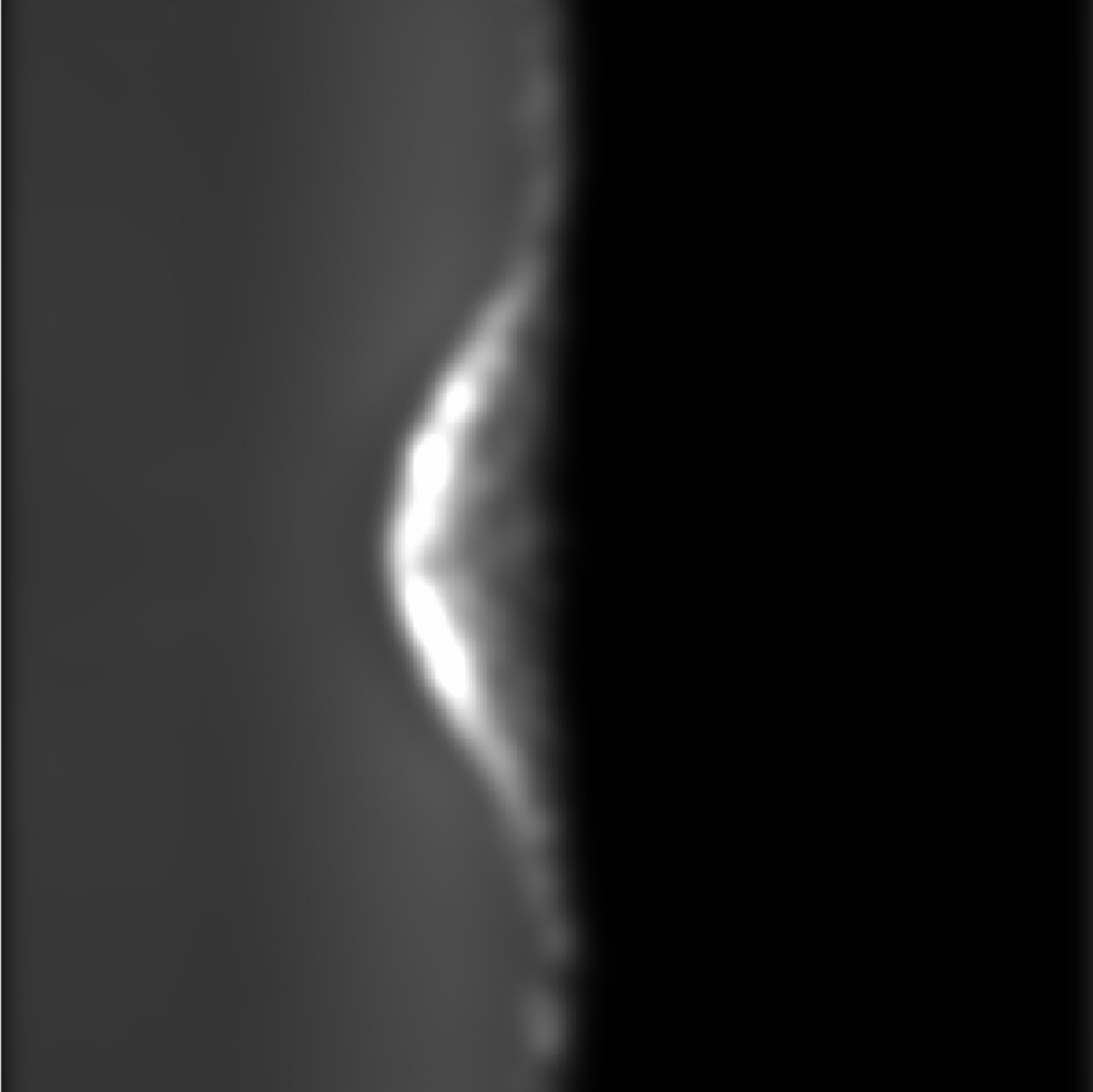}
	\includegraphics[width=4.2cm]{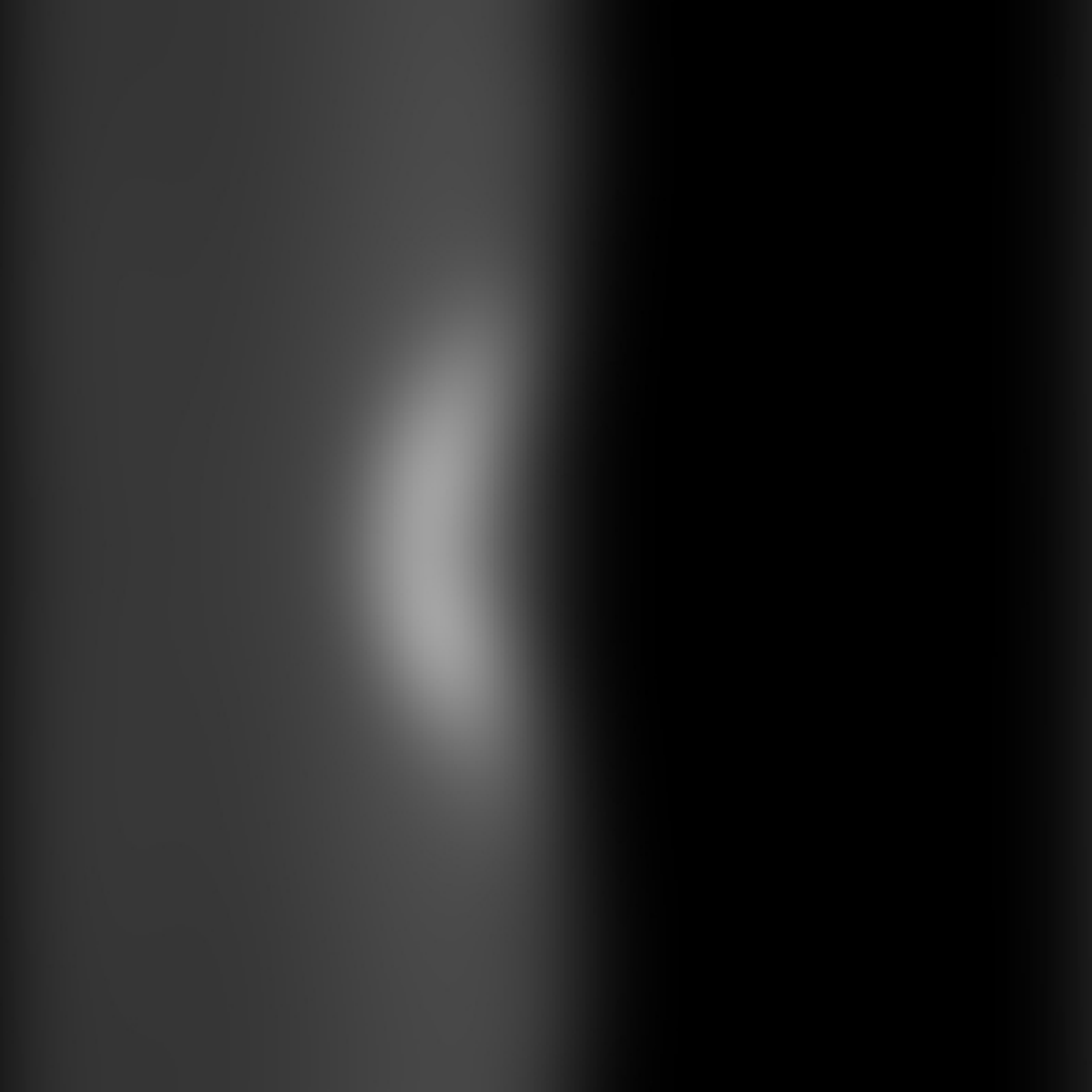}

	\vspace{2pt}
	\includegraphics[width=4.2cm]{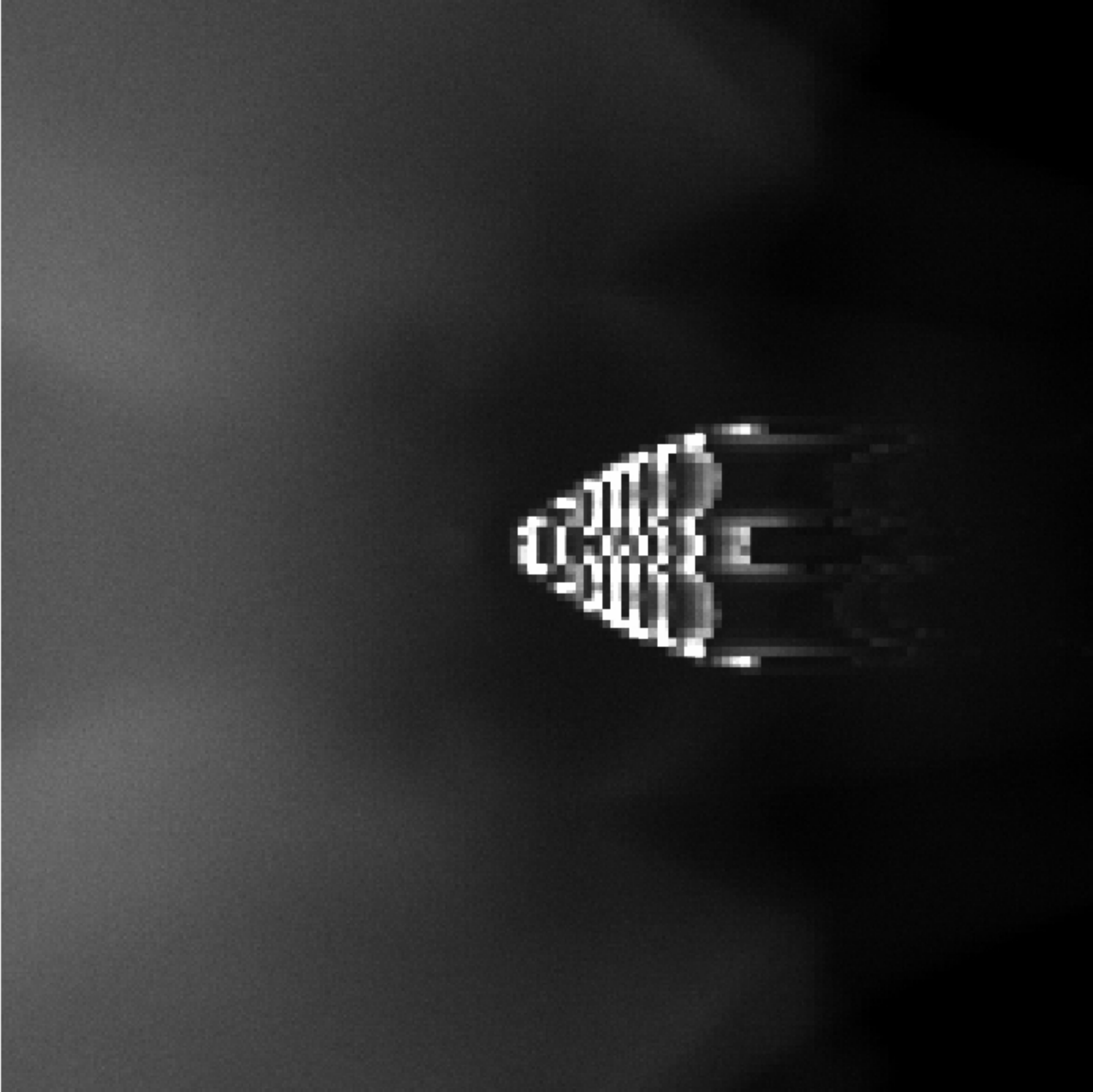}
	\includegraphics[width=4.2cm]{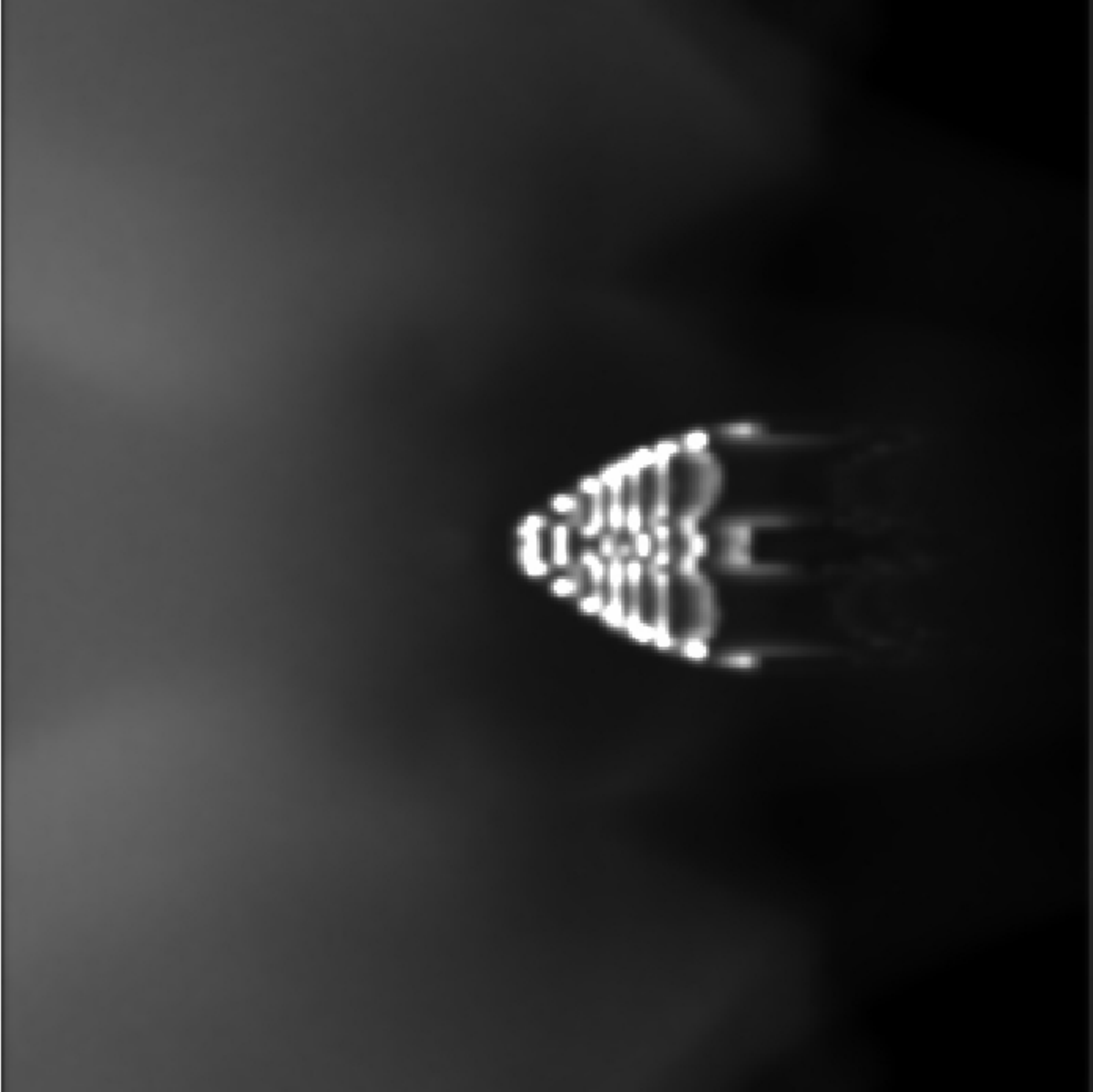}
	\includegraphics[width=4.2cm]{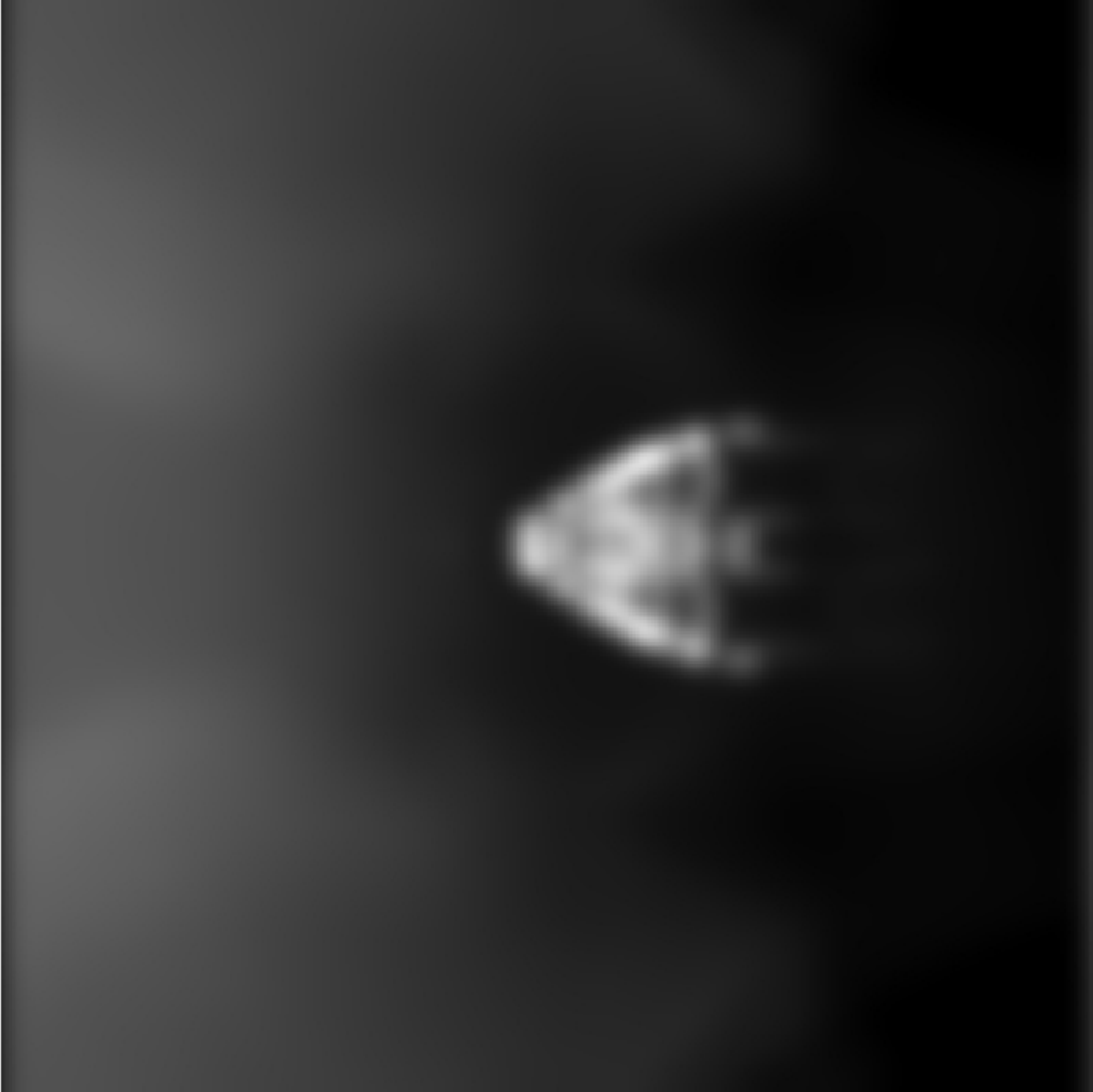}
	\includegraphics[width=4.2cm]{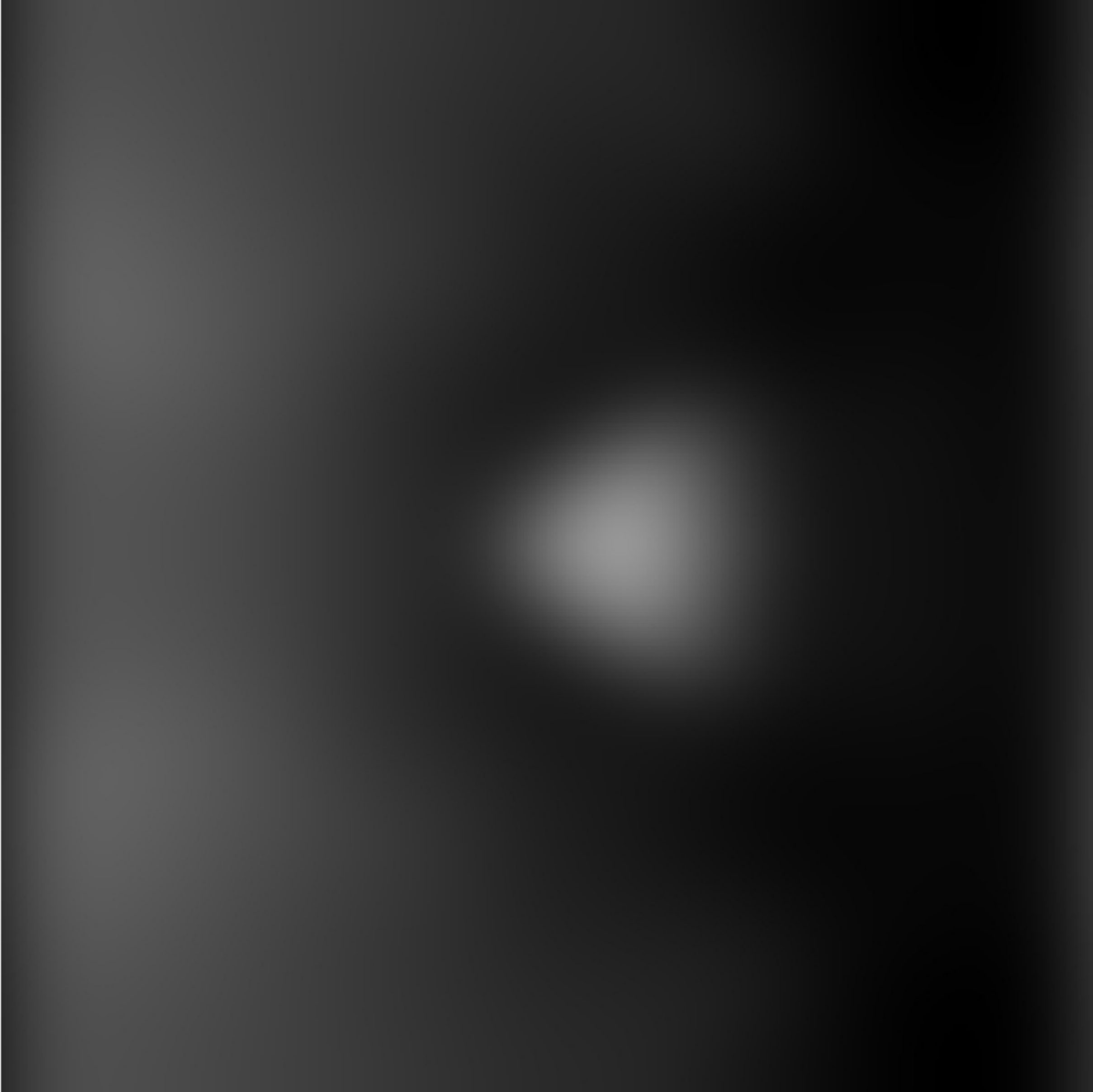}

	\vspace{2pt}
	\includegraphics[width=4.2cm]{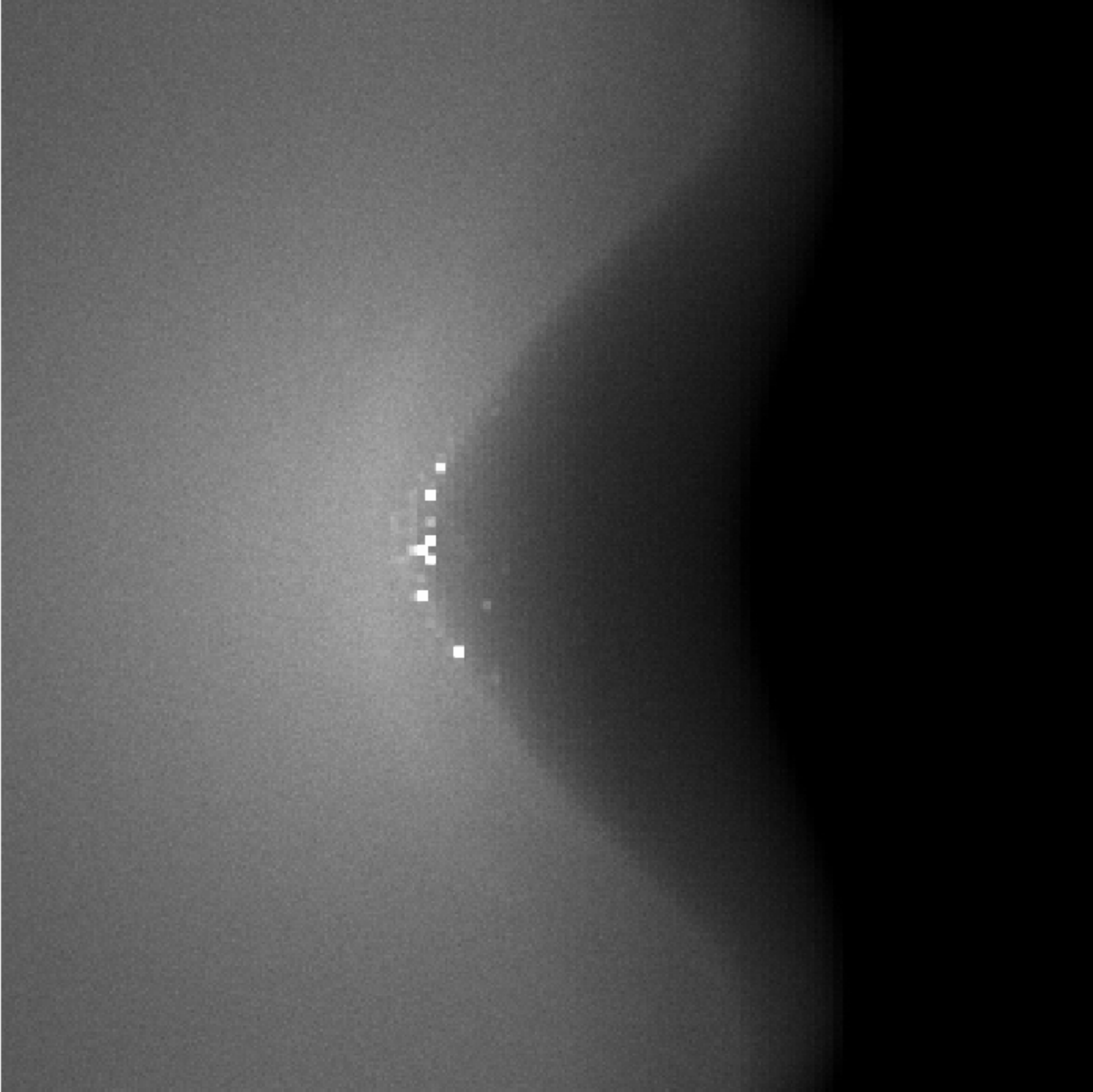}
	\includegraphics[width=4.2cm]{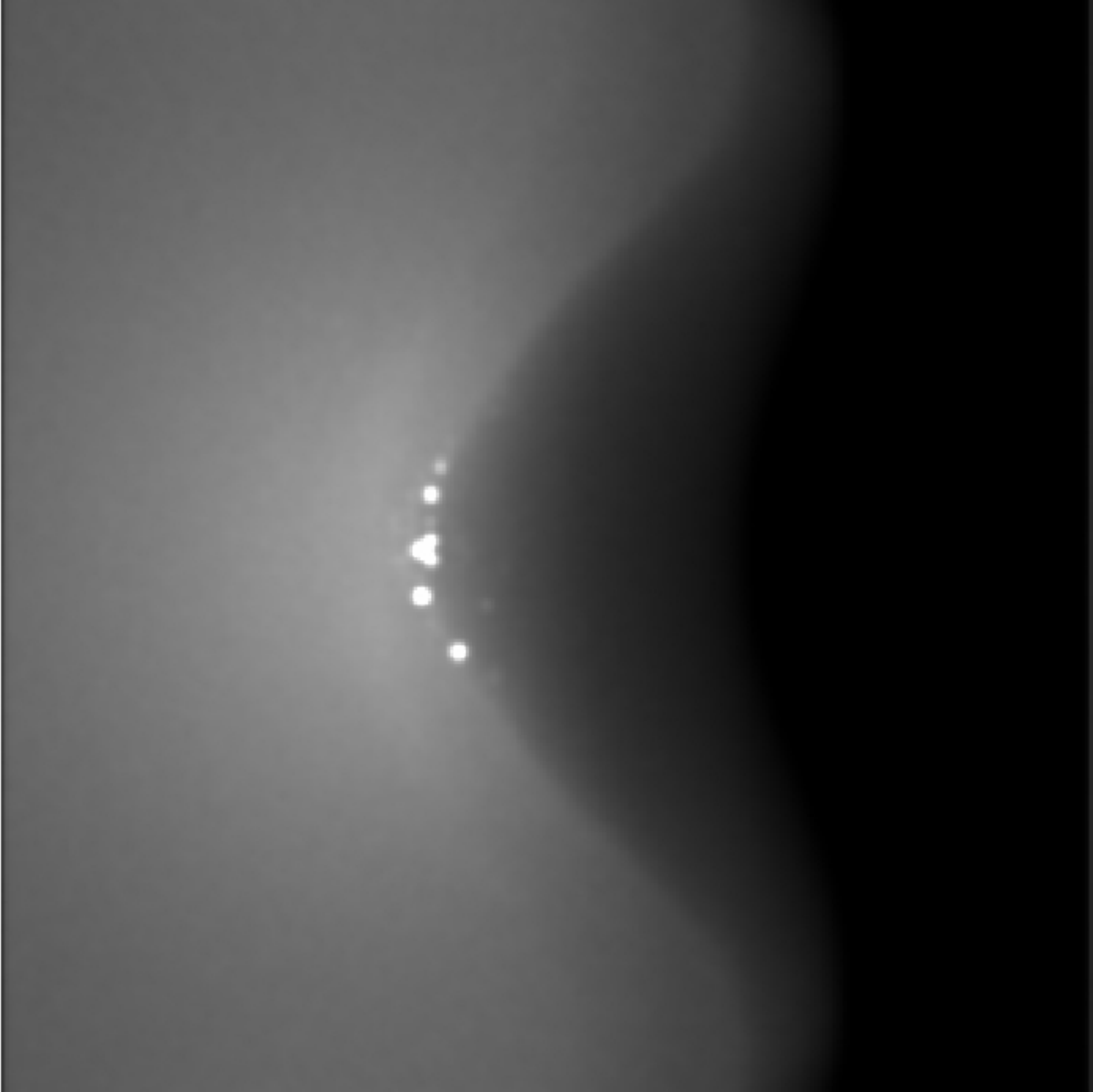}
	\includegraphics[width=4.2cm]{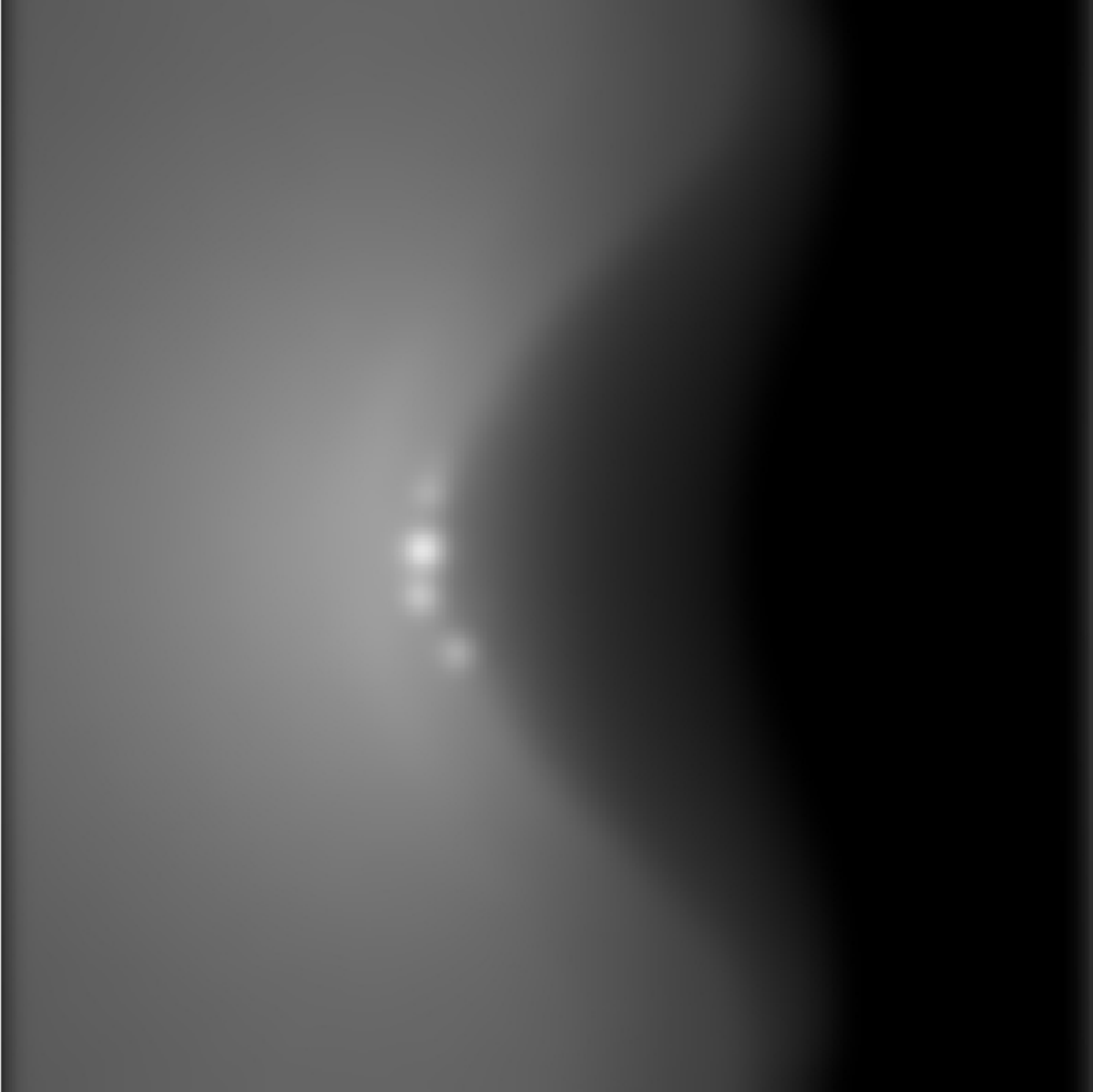}
	\includegraphics[width=4.2cm]{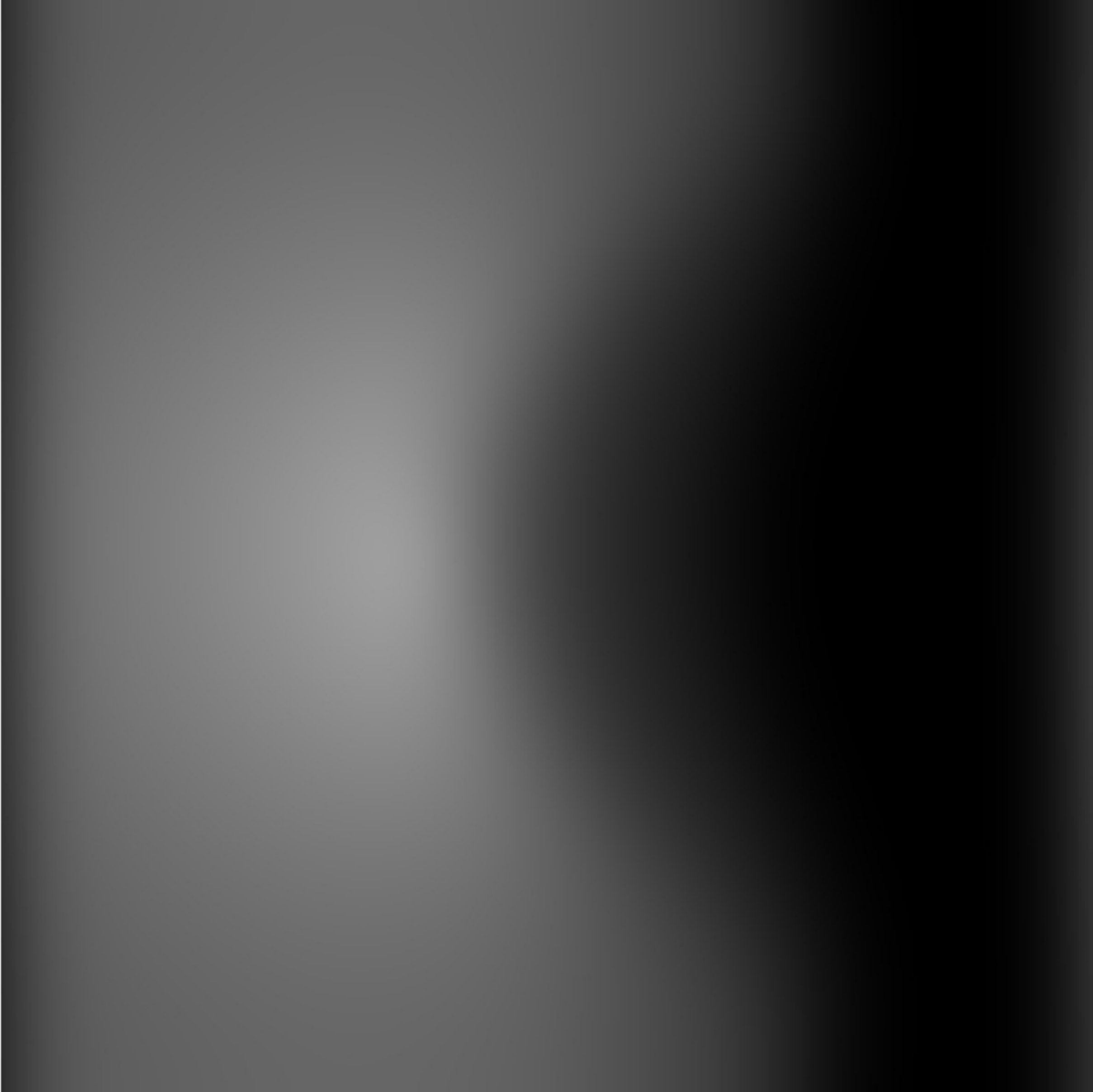}
	\caption{Radio images of the low, medium and high flux models from top to bottom. The left hand panels are the raw output from \textsc{torus}. The other panels, from left to right, are the raw image modified to be respresentative of 30\,s exposures using the VLA type B, C and D configurations. The grey scale of each image is set to match that of the raw output image. Each image side spans 4.87\,pc}
	\label{radioImgs}
\end{figure*}

The raw radio images, as well as simulated  B, C and D configuration, 30 second VLA images are given in Figure \ref{radioImgs}. The colour scale used in each frame is set to match that of the raw image. As the beam size increases, the BRC structure is smoothed out and the brightness of the object is also modified.
The calculated ionizing fluxes, electron densities and mass loss rates based on each radio image are summarised in Table \ref{radioValues}. 
For each model, using a larger beam reduces the measured flux resulting in underestimates of the cloud properties by up to 25\% relative to the unsmoothed image. This is because the compact, bright object flux is partially lost to the surroundings, which are uniformly much dimmer than the BRC.

The ionizing fluxes at the left hand edge of the computational grid are known to be, from low to high flux, $9\times10^8$, $4.5\times10^9$ and $9\times10^9$\,cm$^{-2}$\,s$^{-1}$. The tabulated values of ionizing flux at the BRC, which are calculated at distance of around 2\,pc from the left hand edge of the grid, are therefore of realistic magnitude. There is a discrepancy in that the high flux model ionizing fluxes are only slightly larger than the medium flux, however this is due to the change in ionization structure mentioned in section \ref{tempGrids} that gives rise to an absence of a clear IBL and hence a lower measured flux.

The electron density in the medium flux case has been correctly inferred as being significantly higher than that in the low flux case. The actual number densities in the IBL are in the range $60-190$, $200-1500$ and $50-100$\,cm$^{-3}$ for the low, medium and high flux models respectively. In the medium and low flux cases the calculated values lie beneath this range, suggesting that the effect of the IBL on compression might be underestimated. This underestimate arises due to cooler gas in the regions over which the flux is being integrated at the interior of the bright rim. These relatively cool regions arise where there is localised shielding from the stellar radiation field and radiative heating occurs primarily from the diffuse field. This is the cause of the ribbing effect seen in the medium flux images, where the bright contours are those cells directly exposed to the stellar radiation field. Increasing the region over which the flux is integrated amplifies this problem. For example, in the extreme case of integrating the flux over the entire cloud (region A in Figure \ref{ha_slits}) the low flux electron density is reduced to around 15\,cm$^{-3}$. Conversely, reducing the size over which the flux is integrated too much will make the derived values more susceptible to noise. 

The inferred mass loss rates are higher for the larger clouds, which have a larger surface over which material can be lost. Providing an actual mass loss rate from the computational grids for comparison is non-trivial, particularly in the low flux case where the boundary between the BRC cloud and the gas in the wings of the model is poorly defined. We therefore calculate the difference between the mass at 195 and 200\,kyr in a number of volumes encapsulating the BRCs on the computational grid to provide a range of mass loss estimates. These estimates are also shown in Table \ref{radioValues}, the high flux model is neglected due to the changes to the ionization structure discussed in section \ref{tempGrids}. Both the low and medium flux mass loss estimates from the grid span just over one order magnitude, encompassing the mass loss rates from the 20\,cm emission analysis. 

Although in the correct (but broad) range, the mass loss rate as a quantity does not give a proper indication of the relative strengths of the photo-evaporative flows. For example, the medium flux model ejections are clearly more energetic than the low flux ones, giving rise to stronger rocket-motion, more rapid compression of the cloud and leaving a stronger signature in the ambient HII region, but its mass loss rate is similar. A more useful parameter for studying relative photo-evaporative flow strengths is the mass flux
\begin{equation}
	M_{\rm{f}} = \frac{1}{\Omega}\frac{\dot{M}}{R^2}
\end{equation}
where $R$ is the radius of the cloud and $\Omega$ is the solid angle on the cloud bounded by the IBL. For near edge-on inclinations $\Omega$ can be approximated assuming cylindrical symmetry.  Example mass fluxes are also included in Table \ref{radioValues} which (based on the opening angle of the bow) assume that the low flux IBL bounds $2\pi(1-\cos(\pi/3))$ of the cloud surface and the medium flux cloud IBL bounds $2\pi$ of the surface. These mass fluxes imply that the medium flux flow is significantly stronger than that of the low flux model, in agreement with the behaviour from the model grid. The difficulty in using the mass flux as an indicator of relative photo-evaporative flow strengths is determining what fraction of the cloud is covered by the IBL. A first order comparison of the relative mass fluxes of clouds of similar type could be obtained by assuming that the fraction of the cloud bounded by the IBL is the same between types \citep{1991ApJS...77...59S} and that only the radius of the cloud varies.

\begin{table*}
\centering
  \caption{BRC IBL properties from analysis of radio images at 20\,cm. The mass loss rate from the model grid is also included.}
  \label{radioValues}
  \begin{tabular}{@{}l c c c c c c@{}}
  \hline
   Model & Configuration & Ionizing flux  & Electron density  & Mass loss rate  & Model mass loss rate  & Mass flux\\
   & & ($\Phi$, cm$^{-2}$\,s$^{-1}$) & ($n_{\rm{e}}$, cm$^{-3}$) & ($\dot{M}$, M$_{\odot}$\,kyr$^{-1}$) & ($\dot{M}$, M$_{\odot}$\,kyr$^{-1}$) &  ($M_{\rm{f}}$,\,M$_{\odot}$\,Myr$^{-1}$\,pc$^{-2}$)\\
 \hline
  Low flux & raw & $5.0\times10^8$ &45 & $0.18$ & $1.4\times10^{-2} - 0.2$ & 26\\
  Low flux & B & $4.9\times10^8$ &45 & $0.18$ & $	1.4\times10^{-2} - 0.2$ & 25\\
  Low flux & C & $4.6\times10^8$ &43 & $0.17$ & $1.4\times10^{-2} - 0.2$ & 24\\
  Low flux & D & $3.3\times10^8$ &36 & $0.15$ & $1.4\times10^{-2} - 0.2$ & 21\\
  Medium flux & raw & $1.5\times10^9$ & 122 & $7.8\times10^{-2}$ & $9.8\times10^{-3} - 0.1$ & 35 \\
  Medium flux & B & $1.4\times10^9$ & 119 & $7.7\times10^{-2}$ & $9.8\times10^{-3} - 0.1$ & 34 \\
  Medium flux & C & $1.3\times10^9$ & 113 & $7.3\times10^{-2}$ & $9.8\times10^{-3} - 0.1$ & 32 \\
  Medium flux & D & $8.2\times10^8$ & 91 & $5.9\times10^{-2}$ & $9.8\times10^{-3} - 0.1$ & 26 \\
  High flux & raw & $1.7\times10^9$ & 89 & 0.27 & $-$ & $-$\\
  High flux & B & $1.7\times10^9$ & 89 & 0.27 & $-$ & $-$\\
  High flux & C & $1.6\times10^9$ & 87& 0.26 & $-$ & $-$\\
  High flux & D & $1.4\times10^9$ & 80 & 0.24 & $-$ & $-$\\
\hline
\end{tabular}
\end{table*}

\subsection{Diagnostic line ratio analysis}
\label{ratioAnalysis}
\begin{figure}
	\includegraphics[width=7.84cm]{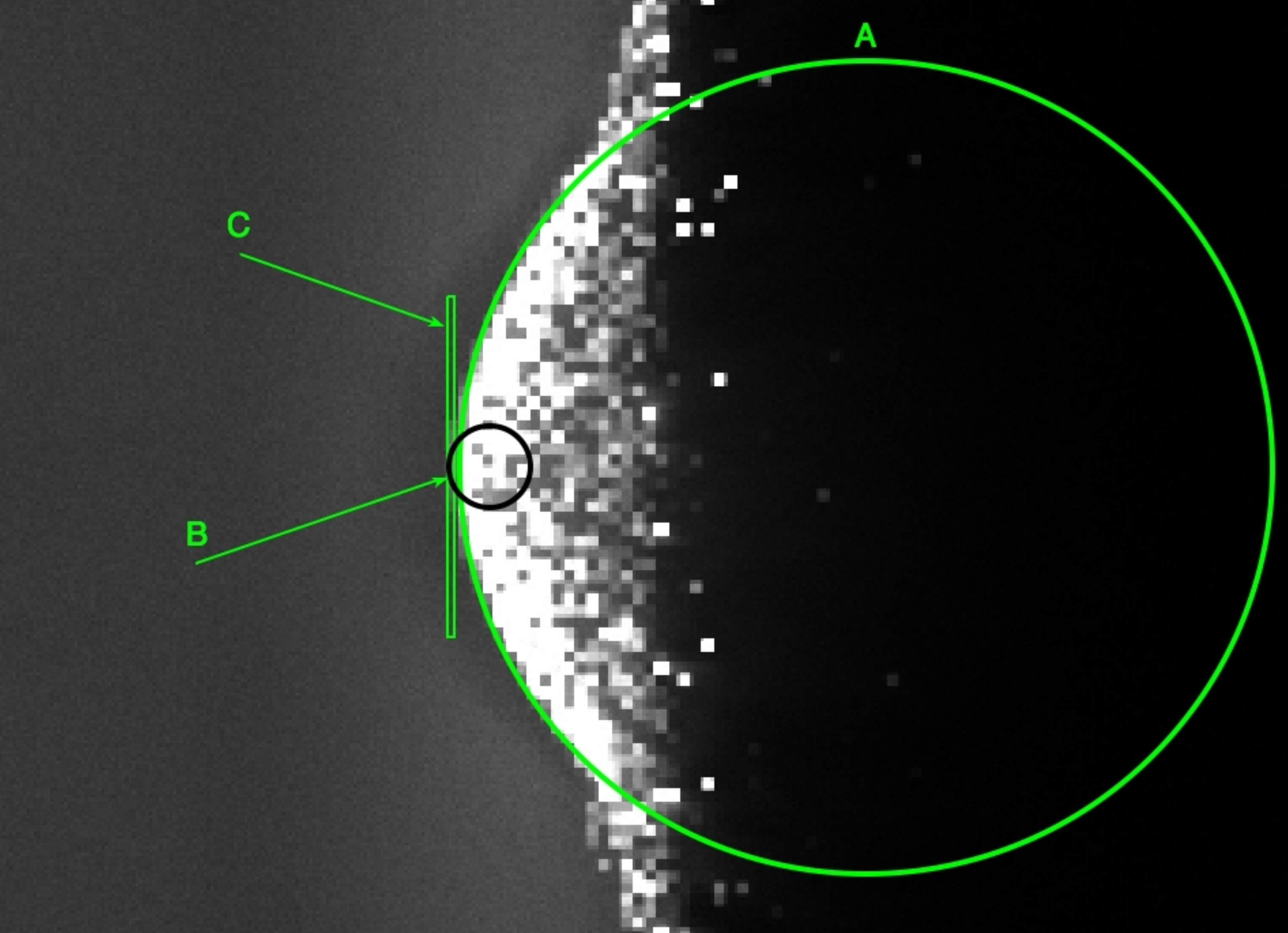}

	\vspace{2pt}
	\includegraphics[width=7.84cm]{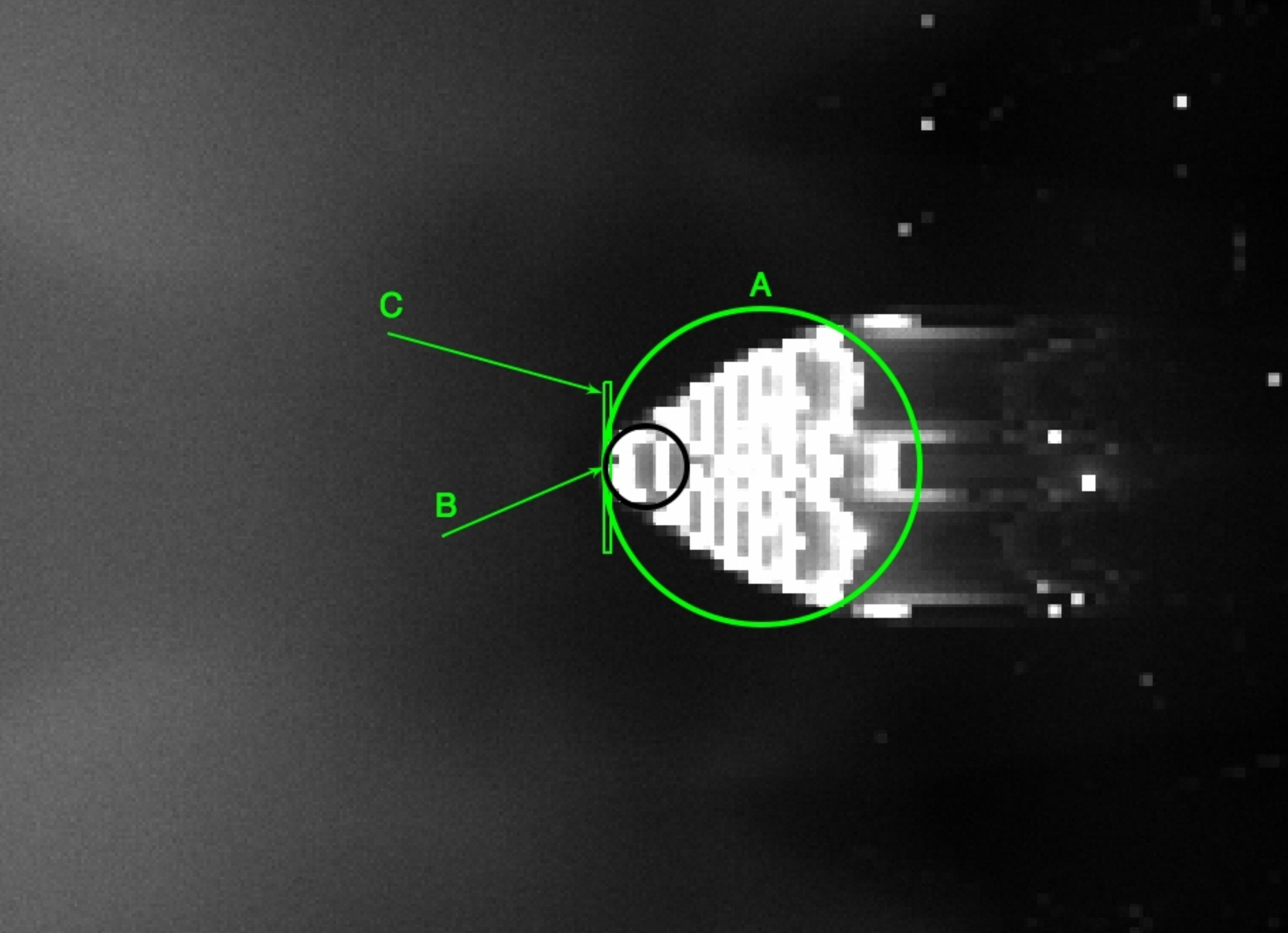}

	\vspace{2pt}
	\includegraphics[width=7.84cm]{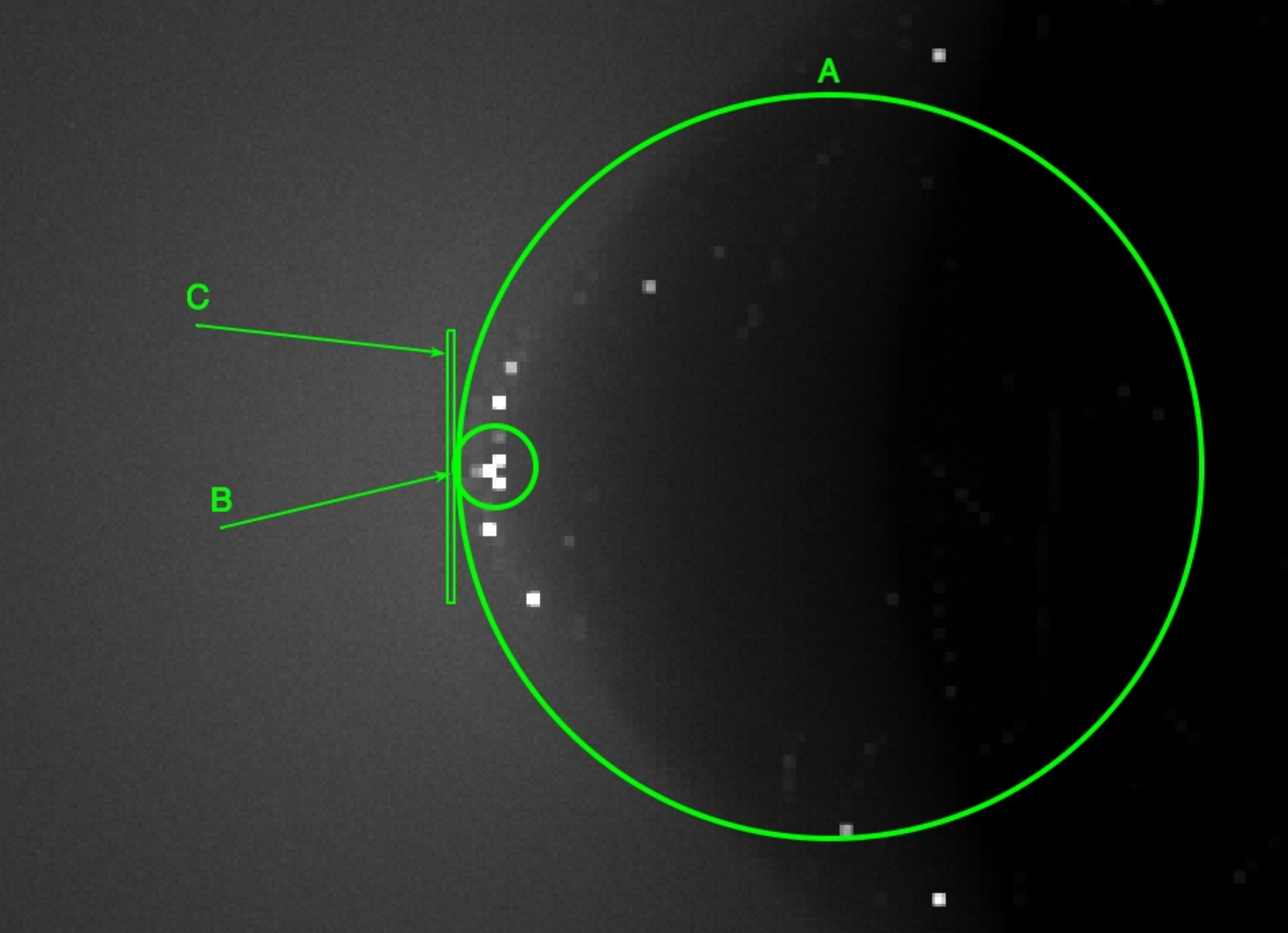}
	\caption{H$\alpha$ images of each model, low to high flux from top to bottom. Overlaid are the regions which are designated as the cloud cross section (those labelled A), the regions over which the radio flux is integrated (those labelled B) and the spectroscopic slits (those labelled C).}
	\label{ha_slits}
\end{figure}

As discussed in section \ref{ratdesc}, we emulate slit spectroscopy by calculating images at specific wavelengths and derive the conditions in the slit region using diagnostic line ratios. The slits are shown in the upper left of H$\alpha$ images of the models in Figure \ref{ha_slits}  (marked C). Each slit has a width of 2 pixels and a length of 100, 50 and 80 pixels for the low, medium and high flux models respectively. These are placed over the ionized cells closest to the tip of the BRC. 

The electron densities and the average of the temperature diagnostics are given in Table \ref{OIInes} along with the values calculated using radio diagnostics and those from the model grids. We retain the level of accuracy determined in section \ref{lextestsec}. As with the radio diagnostic, the electron density is clearly underestimated in the medium and low flux cases (e.g. hydrogen number densities are of order $60-190$\,cm$^{-3}$ for the low flux model). These values could be influenced by the low density (of order ten hydrogen atoms per cubic centimetre) region just beyond the BRC that has been excavated by the photo-evaporative flow. 
\begin{table*}
\centering
  \caption{Diagnostic line ratio electron densities and average IBL temperatures. The average electron densities calculated using the radio diagnostic are also included as well as the known conditions from the model.}
  \label{OIInes}
  \begin{tabular}{@{}l c c c c c@{}}
  \hline
   Model &  Line ratio $n_{\rm{e}}$ (cm$^{-3}$) & Radio $n_{\rm{e}}$ (cm$^{-3}$) & Model $n_{\rm{e}}$ (cm$^{-3}$) & Line ratio $T_{\rm{e}}$ (K)  & Model $T_{\rm{e}}$ (K)\\
 \hline
 Low flux &  20 & $36-45$ & $60-190$ & 8500 & $8000-9000$\\
 & & & \\
 Medium flux &  40 & $91-122$ & $200-1500$ & 8500 & $8000-9000$\\
 & & & \\
 High flux &  60 & $80-91$ & $50-100$  & 8380 & $8000-9000$\\
\hline
\end{tabular}
\end{table*}

The temperatures derived using the [O III], [N II] and [Ne III] ratios are also given in Table \ref{OIInes}. The temperature calculations use the electron densities calculated from the [O II] lines, but depend only weakly on this value so the largest error in $n_{\rm{e}}$ ($\Delta n_{\rm{e}} = 1460$ in the medium flux case) translates into an error of only 15\,K. The separate temperature diagnostics of each model are within 110\,K of one another and all lie within the range seen in the temperature maps of Figure \ref{temperatures}. The actual temperature in the IBL is typically 8000-9000\,K so the average temperatures in Table \ref{OIInes} are more accurate than assuming a value to $10^4$\,K. Given that the electron densities are underestimated by both radio and diagnostic line ratio techniques and the inferred IBL temperatures are more accurate using diagnostic line ratios, this technique is a viable tool for the study of RDI. 

\subsection{Identifying RDI}
\label{idRDI}
If the BRCs are in pressure equilibrium with the IBL then the virial theorem gives
\begin{equation}
	P_{\rm{i}} = \frac{3c_{\rm{s}}^2 M_{\rm{c}}}{4\pi R_{\rm{c}}^3} - \frac{3GM_{\rm{c}}}{20\pi R_{\rm{c}}^4}
	\label{virialA}
\end{equation}
where $M_{\rm{c}}$, $R_{\rm{c}}$, $c_{\rm{s}}$ and $P_{\rm{i}}$ are the cloud mass, cloud radius, cloud gas sound speed and  pressure in the IBL respectively \citep[e.g.][]{2009apsf.book.....H,2004A&A...414.1017T}. If the BRC is not in pressure equilibrium then the contracting or expanding nature of the cloud can be inferred from the relative magnitudes of the left and right hand sides of equation \ref{virialA}, which will be referred to as the external and supporting pressures ($P_{\rm{i}}$ and $P_{\rm{s}})$ respectively. 

For the radio method we use the electron densities calculated in section \ref{radioAnalysis} and make the standard assumption that the ionized gas is at 10000\,K. For the diagnostic ratio method we use the electron densities and the average temperatures calculated in section \ref{ratioAnalysis}. The neutral cloud properties are those determined in section \ref{greybodyAnalysis} using $C_{\nu}$=214\,g\,cm$^{-2}$. Under this virial equilibrium technique a cloud radius needs to be assumed, the circular regions on Figure \ref{ha_slits} represent the cross section of the spherical clouds used to investigate equilibrium. 

The low, medium and high flux clouds have radii of 1.5, 0.6 and 1.3\,pc respectively. Given that most of the neutral mass on the grid is concentrated in the BRC, we assume in each stability analysis that all of the SED fitted mass is contained within the highlighted regions. The supporting cloud pressures for each model, from low to high flux are $4.6\times10^{-13}$, $4.1\times10^{-12}$ and $3.7\times10^{-13}$\,dyn\,cm$^{-2}$ respectively. Typical pressures in the neutral cloud on the radiation hydrodynamic simulation grid are $10^{-13}-10^{-12}$, $10^{-12}-10^{-10}$ and $10^{-13}-10^{-12}$\,dyn\,cm$^{-2}$ in the low, medium and high flux models respectively so the derived pressure values are towards the centre of this range in the low and high flux models and towards the lower end of the range found in the medium flux model.

%underestimates of the isothermal cloud support pressure. This is an unsurprising consequence of underestimating the neutral cloud masses as discussed in section \ref{greybodyAnalysis}.

The external pressures and their ratio relative to the support pressure are given in Table \ref{virialtable}. 
The pressures in the IBL from the model grids range from around $0.5-3\times10^{-10}$, $1-5\times10^{-10}$ and $1-2.5\times10^{-10}$\,dyn\,cm$^{-2}$ for the low, medium and high flux models respectively. The IBL pressures derived using diagnostic line ratios lie within this range, however because the assumed temperature of $10^4$\,K is larger than the actual temperature in the IBL the radio diagnostics overestimate the pressure in the IBL in the medium and high flux cases. The low flux electron density was sufficiently underestimated for the pressure to fall into the range on the model grid.

\begin{table*}
\centering
  \caption{IBL pressures and their ratio to the supporting cloud pressure in virial stability analysis. The ratio of IBL to neutral cloud supporting pressures from the model are also included.}
  \label{virialtable}
  \begin{tabular}{@{}l c c c c@{}}
  \hline
   Model & Diagnostic method & External pressure ($P_{\rm{i}}$, $\times10^{-10}$\,dyn\,cm$^{-2}$) & $P_{\rm{i}}/P_{\rm{s}}$  & Model  $P_{\rm{i}}/P_{\rm{s}}$\\
 \hline
  Low flux & Radio, raw image &  2.1 & 460 & 50-3000\\
  Low flux & Radio, VLA configuration B &  2.1 & 460 & $50-3000$\\
  Low flux & Radio, VLA configuration C &  2.0 & 430 & $50-3000$\\
  Low flux & Radio, VLA configuration D &  1.7 & 370 & $50-3000$\\
  Low flux & Slit spectroscopy & 0.78 & 170 & $50-3000$\\
  Medium flux & Radio, raw image &  5.6 & 140 & $1-500$ \\
  Medium flux & Radio, VLA configuration B & 5.5  & 130 & $1-500$\\
  Medium flux & Radio, VLA configuration C &  5.2 & 130 & $1-500$\\
  Medium flux & Radio, VLA configuration D &  4.2 & 100 & $1-500$\\
  Medium flux & Slit spectroscopy &  1.6 & 40 & $1-500$\\  
  High flux & Radio, raw image &  4.1 & 1110 & $100-2500$\\
  High flux & Radio, VLA configuration B &  4.1 & 1110 & $100-2500$\\
  High flux & Radio, VLA configuration C &  4.0 & 1080 & $100-2500$\\
  High flux & Radio, VLA configuration D &  3.7 & 1000 & $100-2500$\\
  High flux & Slit spectroscopy &  2.3 & 620 & $100-2500$\\
\hline
\end{tabular}
\end{table*}
All diagnostics imply that the cloud should be undergoing thermal compression, which is in qualitative agreement with the known model behaviour in the medium and low flux cases. Furthermore, the difference in pressure is weaker in the medium flux case than the low flux case. This reflects the relative behaviours of the clouds, as the rate of compression of the medium flux cloud has started to plateau by this point (200\,kyr) in the radiation hydrodynamics calculation. The biggest pressure differences are found in the high flux model. Although this is not in agreement with the pressure supported system at the time of imaging in the radiation hydrodynamics calculation, the ionization front has been relocated to a point at which D-type driving of the front into the cloud is about to occur as discussed in section \ref{tempGrids}. 
Given these modifications to the starting conditions, it is likely that the resulting BRC following a new calculation with metals would more closely resemble that of the medium flux model. 
The ratio of external to supporting pressure lies within the range of those found on the computational grid in all cases, but are typically towards the lower end of the range. This is due to a combination of the overestimated neutral cloud masses and the underestimated IBL electron densities.

\section{Summary and conclusions}
We have performed a range of synthetic observations and diagnostics of radiation hydrodynamic RDI model results to investigate the accuracy of diagnostic techniques and to identify signatures of RDI. We have produced SEDs, images that are representative of 30 second VLA radio observations using the type B, C and D configurations (the latter of which is used for the NRAO VLA Sky Survey) and forbidden line images of each model. These synthetic observations have been used to replicate a number of diagnostics to calculate the conditions in the neutral BRC and its IBL. Using these conditions we have performed a virial stability analysis of each system to determine whether the BRCs are being compressed by the IBL.
 
We draw the following conclusions from our imaging and diagnostics:
\\

1. The synthetic images generated show objects which are morphologically similar to BRCs observed in star forming regions. 
\\

2. The neutral cloud dust temperatures derived using greybody fitting of the SED are similar to those calculated in observational studies \citep[e.g.][]{2004A&A...414.1017T, 2008A&A...477..557M} and are slightly higher than, but within 2\,K of, the known dominant cloud temperature (10\,K for at least $95\%$ of the cloud by mass). This slight temperature overestimate is due to warmer dust in the cloud. The mass inferred using equation 10a from \cite{1983QJRAS..24..267H} (equation \ref{dustGasMass} in this paper) at 10\,K is found to vary by $8-16$\,\% with a change of 1\,K and by up to 35\,\% with a change of 2\,K. Using a fixed value of $C_{\nu}$ for clouds of different class is found to induce an error in the calculated mass of up to a factor 3.6.
\\

3. The temperatures derived using diagnostic line ratios at around 8500\,K are more accurate than assuming a canonical temperature of $10^4$\,K as in the radio diagnostics. The electron number densities established using both techniques are underestimates of those in the IBL on the computational grid. Diagnostic line ratios underestimate the electron density due to contamination from the low density region excavated by the photo-evaporative flow. The radio electron density underestimate is due to lower emission from the interior regions of the BRC where some shielding from the stellar radiation field occurs and radiative heating is due to the, relatively weak, diffuse field.
\\

4. The supporting cloud pressures are found to lie within the range on the computational grids for all models. The IBL pressures are found to be within the correct range for the forbidden line ratio diagnostic. However, since the assumed temperature in the IBL is larger than the actual temperature the radio diagnostics overestimate the IBL pressure in the medium and high flux cases. The IBL pressure is greater than that of the supporting cloud for each model, implying that the cloud is being compressed. This behaviour is qualitatively correct for the low and medium flux models, however is not the observed behaviour of the pressure balanced high flux model. The ratio of external to supporting pressure is found to be towards the lower end of the range of values found on the model grid in each case.
\\

5. The ionization state of the low and medium flux grids following the initial photoionization calculation remained reasonably consistent with that at the end of the radiation hydrodynamic calculations. However, for the high flux model in which the density structure in the system was not significantly altered in the radiation hydrodynamics calculation, the ionization front has moved closer to the star. This is due to the enhanced cooling by forbidden line emission and more comprehensive thermal balance calculation, essentially resetting the model to a point in which D-type expansion of the ionization front is yet to occur. With a greater distance over which to accumulate material, the modified high flux system may well have achieved a stronger photo-evaporative flow and higher levels of compression of the BRC than in the original model. This result indicates the importance of including atomic chemistry in future calculations. 
\\

6. The effect of moving to larger beam sizes in radio observations is to contaminate the integrated flux from the BRC with that from the surrounding neutral cloud and HII region. Since both the neutral cloud and HII region are dimmer than the IBL this reduces the integrated flux, resulting in an underestimate of the incident ionizing flux, electron density and mass loss rate relative to that calculated using an unsmoothed comparison image
\\

7. Despite the presence of strong photo-evaporative flows which establish prominent, relatively cool, low density regions on the model grid, they are difficult to detect in the images calculated here. No striations are detected in H\,$\alpha$ about the BRC, rather the visual indicator of photo-evaporative flow in these systems is darker regions in the vicinity of the BRC, where material is excavated by the flow. It is likely that striations are not observed because the resolution in these calculations was not sufficiently high to resolve them. 
\\

8. The estimate for mass loss rate of clouds from \cite{1994A&A...289..559L} is found to be in good agreement with that of the clouds on the computational grid. By using the mass flux, a measure of the relative strengths of the photo-evaporative flows can be calculated, correctly implying that the medium flux model exhibits much stronger flow than the low flux model, despite having a similar mass loss rate. 
\\

We have shown that existing diagnostics do give an insight into the conditions of BRCs and can broadly be used to infer whether or not RDI is occuring using virial stability analysis. However, by comparing the inferred conditions to those on the model grids we find that these diagnostics are each subject to significant sources of error. The cumulative effect of these errors is IBL-to-cloud pressure ratios towards the lower end of the known range, suggesting that the effects of shock compression might be underestimated.

We next intend to perform synthetic diagnostics using molecular lines, which are a useful tool for investigating the kinematic behaviour of astrophysical gases \citep[e.g.][]{2002ApJ...577..798D,2006A&A...450..625U,2009MNRAS.400.1726M,2010MNRAS.407..986R}. Applying these techniques we can draw comparison conculsions to those in this paper from a more sophisticated analysis of the gas, investigate the kinematic behaviour of BRCs and provide a test of widely used molecular line diagnostics.

\section*{Acknowledgments}
The calculations presented here were performed using the University of
Exeter Supercomputer, part of the DiRAC Facility jointly funded by
STFC, the Large Facilities Capital Fund of BIS, and the University of
Exeter. T. J. Haworth is funded by an STFC studentship.  
The authors thank Mark Thompson and Chris Brunt for useful discussions and the anonymous referee for their useful comments.

\bibliographystyle{mn2e}
\bibliography{synthetics}

\appendix

\bsp

\label{lastpage}

\end{document}